\documentclass[sigconf]{acmart}

\copyrightyear{2018} 
\acmYear{2018} 
\setcopyright{acmcopyright}
\acmConference[SIGMOD'18]{2018 International Conference on Management of Data}{June 10--15, 2018}{Houston, TX, USA}
\acmPrice{15.00}
\acmDOI{http://dx.doi.org/10.1145/XXXXXX.XXXXXX}
\acmISBN{ISBN 978-1-4503-4703-7/18/06} 

\usepackage{booktabs} 
\setcopyright{none}

\usepackage{url}
\usepackage{amsmath}
\usepackage{amssymb}
\usepackage{color}
\usepackage{xspace}
\usepackage{algpseudocode}
\usepackage{algorithm}
\usepackage{array}
\usepackage{subcaption}
\usepackage{enumerate}
\usepackage{graphicx}
\usepackage{balance}
\usepackage{soul}
\usepackage{paralist}
\usepackage{fancyvrb}

\newcommand{\cut}[1]{}
\newcommand{\ads}[1]{}

\newtheorem{assumption}{Assumption}
\newtheorem*{assumption*}{Assumption}
\newtheorem{problem}{Problem}
\newtheorem{example}{Example}

\newcommand{\algname}{{\sc Datamaran}\xspace}
\newcommand{\scoreF}{{regularity score function}\xspace}
\newcommand{\score}{{regularity score}\xspace}

\newcommand{\apprxscoreF}{{assimilation score function}\xspace}
\newcommand{\apprxscore}{{assimilation score}\xspace}
\newcommand{\stitle}[1]{\vspace{0.5em}\noindent\textbf{#1}}

\newcommand{\techreport}[1]{#1}
\newcommand{\paper}[1]{}

\newcommand{\new}[1]{#1} 

\newenvironment{denselist}{
	\begin{list}{\small{$\bullet$}}%
		{\setlength{\itemsep}{0ex} \setlength{\topsep}{0ex}
			\setlength{\parsep}{0pt} \setlength{\itemindent}{0pt}
			\setlength{\leftmargin}{1.5em}
			\setlength{\partopsep}{0pt}}}%
	{\end{list}}

\begin{document}

\title{Navigating the Data Lake with {\sc Datamaran}: \\ Automatically Extracting Structure from Log Datasets}

\author{Yihan Gao}
\affiliation{%
	\institution{University of Illinois at Urbana-Champaign}
	\city{Urbana} 
	\state{Illinois} 
	\postcode{61801}
}
\email{ygao34@illinois.edu}

\author{Silu Huang}
\affiliation{%
	\institution{University of Illinois at Urbana-Champaign}
	\city{Urbana} 
	\state{Illinois} 
	\postcode{61801}
}
\email{shuang86@illinois.edu}

\author{Aditya Parameswaran}
\affiliation{%
	\institution{University of Illinois at Urbana-Champaign}
	\city{Urbana} 
	\state{Illinois} 
	\postcode{61801}
}
\email{adityagp@illinois.edu}

\begin{abstract}
Organizations routinely accumulate 
semi-structured log datasets generated as the output of code; 
these datasets remain unused and uninterpreted, 
and occupy wasted space---this phenomenon has been colloquially referred to as
{\em ``data lake''} problem. One approach to leverage these semi-structured  
datasets is to convert them into a structured relational format, following
which they can be analyzed in conjunction with other datasets.
We present \algname, an tool that extracts structure from semi-structured 
log datasets with no human supervision. 
\algname 
automatically identifies field and record endpoints, separates 
the structured parts from the unstructured noise or formatting,
and can tease apart multiple structures from within a dataset,
in order to efficiently extract structured relational datasets
from semi-structured log datasets, at scale with high accuracy.
Compared to other unsupervised log dataset extraction tools 
	developed in prior work, \algname does not require the record 
	boundaries to be known beforehand, making it much more applicable to
	the noisy log files that are ubiquitous in data lakes.
\algname can successfully extract structured information from all datasets 
used in prior work, and can achieve $95\%$ extraction accuracy on 
automatically collected log datasets from GitHub---a substantial 66\% increase of accuracy compared to unsupervised schemes from prior work. 
\new{Our user study further demonstrates that the extraction results of \algname are  
closer to the desired structure than competing algorithms.}
\end{abstract}

\maketitle


\section{Introduction}\label{sec:introduction}

Enterprises routinely collect semi-structured or partially
structured log datasets in shared file systems such as HDFS.
These datasets are typically generated automatically
as log datasets output by programs, 
and often number in the billions, e.g., Google 
has 26B datasets in their shared file system~\cite{halevy2016goods}.
This phenomenon of accumulation of log datasets within
enterprises
has recently been referred to as
the {\em ``data lake'' problem}~\cite{stein2014enterprise,terrizzano2015data,hai2016constance}.
Unfortunately, the datasets in a data lake
often remain {\em unused}, {\em unstructured}, and {\em uninterpreted}, and as
they accumulate, they become {\em unmanageable}---recent
work has characterized this data lake problem as one of the most
important challenges facing large enterprises today~\cite{datalake-gartner,stein2014enterprise}.

The first step to making these log datasets more useful is
to convert them into a structured (relational) format.
Once we have structured these datasets, we can then infer relationships
 across datasets, and use them to aid analysis, search, or browsing~\cite{infogather,infogatherplus,DBLP:journals/pvldb/CafarellaHWWZ08,DBLP:journals/pvldb/VenetisHMPSWMW11,DBLP:conf/cloud/GonzalezHJLMSS10,DBLP:conf/sigmod/SarmaFGHLWXY12,DBLP:journals/pvldb/LimayeSC10,chakrabarti2016data}.
The goal of this paper is to \ul{\em automatically, efficiently, and accurately extract structure from log datasets}, 
enabling us to tap into  the log datasets in large enterprise data lakes. 

\stitle{Why Not Use Prior Work?} Given the vast volumes of related work on information extraction~\cite{sarawagi2008information},
one may be tempted to ask: doesn't that solve the problem? 
Unfortunately, as we will describe in more detail in Section~\ref{sec:related},
much related work on general HTML 
wrapper induction, e.g., \cite{dalvi09robust,freitag2000boosted,han01wrapping,hsu98generating,muslea98stalker,vertex},
HTML list-based extraction, e.g., \cite{gupta2009answering,machanavajjhala2011collective}, 
and others, e.g., \cite{senellart2008automatic,li2008regular}, requires training examples
or a corpus of entities to be provided. 
A relatively smaller body of work exists on unsupervised extraction, from
general HTML pages~\cite{arasu2003extracting,sleiman2014trinity,sleiman2013tex}, 
and HTML lists~\cite{elmeleegy2009harvesting,chjwww02,zhai2005web}. 
The former crucially relies on the HTML DOM tree, opting to identify recurrent tree patterns;
and the latter relies on having each list item corresponding to a record. 
Log datasets unfortunately do not correspond to a tree structure
and records in log datasets are often of multiple types, and span multiple lines, making it hard to identify
record boundaries. Moreover, records are interspersed 
noise or other formatting, making it hard to apply the HTML list techniques. 
\new{Finally, unsupervised extraction techniques designed for other media, e.g., network protocols, or natural language corpora~\cite{Cohen2011,Spitkovsky2011,antunes2011reverse,cui2007discoverer},
crucially rely on characteristics of the datasets they are targeting, and are not applicable to log dataset extraction.}

\begin{figure}[h]
	\vspace{-10pt}
	\begin{center}
		\includegraphics[width = 7.5cm]{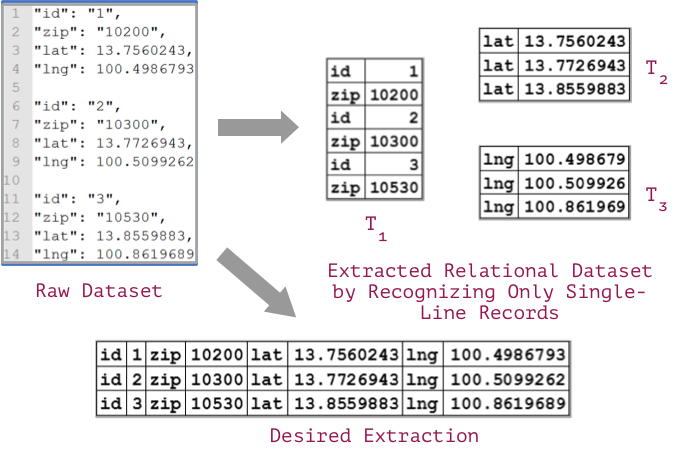}
	\end{center}
	\vspace{-10pt}
	\caption{\new{Sample multi-line record dataset, along with the extraction results of line-by-line extraction schemes.}\label{fig_sl_vs_ml}}
	\vspace{-10pt}
\end{figure}

Perhaps the most related body of work is on log dataset extraction itself.
Work from program synthesis has developed techniques
to perform extraction or transformation 
from examples~\cite{le2014flashextract,DBLP:conf/popl/Gulwani11,DBLP:journals/cacm/GulwaniHS12,jin2017foofah},
while some others~\cite{kandel2011wrangler,raman2001potter} require the users to provide the transformation steps; 
instead, we are opting for a fully unsupervised approach. 
Fisher et al.~\cite{fisher2008dirt} take one step towards automation 
by only requiring that users
provide record boundaries: 
they assume that the data is already {\em chunked} 
(i.e., partitioned into small blocks such that each block contains exactly one record) 
beforehand using external tools. 
This chunking step is assumed to be a simple form of supervision 
(e.g., when each record contains exactly $k$ lines), and 
their work primarily focus on learning structure given the blocks. 
Recordbreaker~\cite{recordbreaker} is a simple automated implementation
of Fisher et al.'s technique that assumes that
each record occupies exactly one line. 
As we will see below, this is far too drastic an assumption to retain applicability in a data lake scenario.

\begin{figure}[h]
	\vspace{-5pt}	
	\begin{center}
		\includegraphics[width = 7cm]{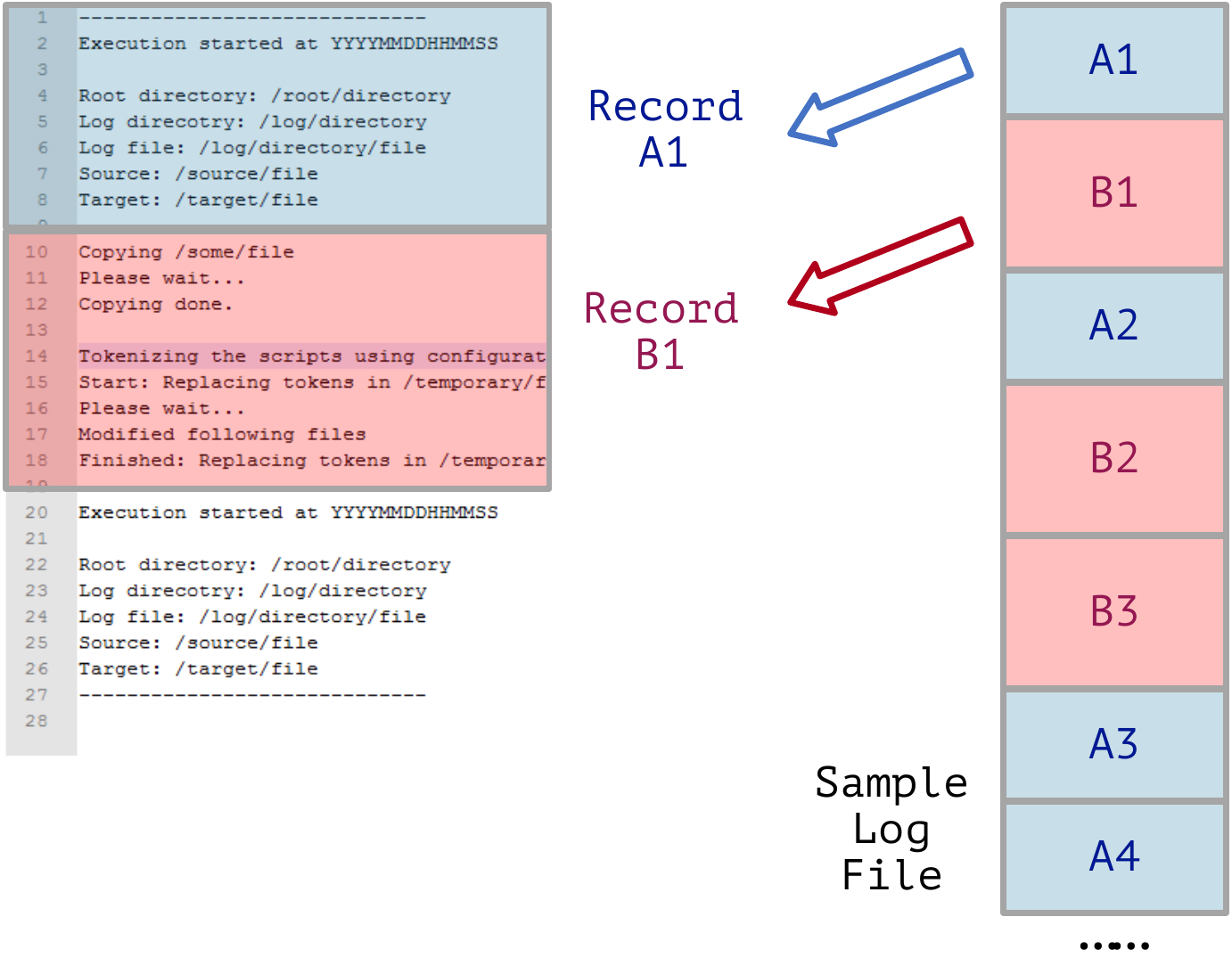}
	\end{center}
	\vspace{-10pt}
	\caption{\new{Sample log dataset from GitHub with contents anonymized; only first few lines are shown.}}\label{fig_sample_file}
	\vspace{-5pt}
\end{figure}

\new{
\begin{example}[Importance of Recognizing Multi-line Records]	
\vspace{-10pt}
Consider the example log dataset\footnote{\scriptsize \new{The example dataset in this figure is a simplified version of the Thailand district info dataset, which is one of the $25$ datasets used in our evaluation in Section~\ref{sec_experiment} and in our user study in Section~\ref{sec_user_study}.}} in Figure~\ref{fig_sl_vs_ml}, where each record occupies multiple lines. 
One promising approach to extract from such a dataset is to use an unsupervised extraction scheme such as RecordBreaker~\cite{recordbreaker} that extracts contents from each line independently, resulting in tables $T_1, T_2, T_3$ as displayed, such that a multi-line record can be viewed to be a union of multiple single-line records. 
While such a line-by-line extraction approach indeed can potentially ``extract'' all relevant content,
the associations between records are completely lost in the generated tables.
The loss of record association information makes it very hard for users to 
interpret or utilize such results, say, for example, for keyword search or data integration, 
both of which require join paths to be preserved. On the other hand, 
our tool, \algname, described next, extracts multi-line records correctly as single records.
\end{example}
}

\new{
\begin{example}[Importance of Recognizing Multiple Record Types]
	Consider an example log dataset crawled from 
	GitHub in Figure~\ref{fig_sample_file}, in which there are two types of records  (A and B) consisting 
	of $7$ and $9$ lines respectively, randomly interspersed with each other. 
	Since the sequence of record types can be arbitrary, 
	it is no longer possible to identify the boundaries of records using simple rules, 
	rendering prior unsupervised log structure extraction algorithms non-applicable\footnote{\scriptsize Note, although that while in this example, the boundaries of records are represented as the special ``--------'' lines, an unsupervised chunker cannot utilize this information without human guidance}.
	\vspace{-5pt}
\end{example}
}


\stitle{\algname: Automatic Log Structure Extraction.} 
In this paper, we present {\sc \algname}\footnote{\scriptsize Catamaran is a type of 
boat or raft meant to rapidly navigate large water bodies, such as lakes or oceans.}, 
an automatic log dataset structure extraction algorithm. 
At a high-level, the idea behind \algname is simple:
\algname identifies the correct structure of the dataset by looking for repeated patterns: 
we examine small portions of the dataset and use a hash-table to find repeated patterns 
covering a significant fraction of the dataset. 
All such patterns are then evaluated via some scoring function, such as the minimum description length~\cite{barron1998minimum} (Note, however, that \algname is general, and can adapt to any scoring modality, not just minimum description length). 
Finally, the best pattern is used to actually extract structured information from the dataset.

However, a naive implementation of this algorithm, as we will demonstrate,
can lead to a huge blowup in the number of patterns considered, and therefore the time
taken for extraction; as a result, \algname requires careful design and engineering
to bound the computation at each step. 
We developed techniques to address the following challenges we encountered when applying the above high-level idea on log datasets:
\begin{denselist}
	\item \emph{Unknown Record Endpoints.} As described above, identifying the boundaries of
	records is not straightforward; while the end-of-line character `\textbackslash n' is 
	often used for separating records, it could also appear within records (i.e., multi-line records).
	\item \emph{Unknown Field Endpoints.} When trying to detect repeated patterns, 
	it is necessary to separate the formatting characters from the field values. 
	This is not as easy in log datasets, 
	due to the fact that commonly used formatting characters (e.g., the space character ` ') 
	can also appear within field values (e.g., text fields). 
	\item \emph{Complex Structure.} There are often complex structures within records: 
	e.g., if a record contains a list of values, the number of values can vary from record to record, 
	which makes even the underlying formatting vary between records, and therefore the same pattern not applying
	across the dataset. 
	Indeed, like our example demonstrates, multiple record types may also exist within the dataset. 
	Furthermore, substructures could also exist within the structures via nesting. 
	This makes detecting repetitive patterns substantially more difficult.
	\item \emph{Redundant Structure.} During the early stages of extraction we often find a number of different repetitive patterns; of which most are completely useless (e.g., the trivial pattern that extracts the entire dataset). The number of such patterns can blow up very quickly as the structure becomes complex: for example, the date component \verb|YYYY-MM-DD| ~can be identified as either a single field or three different fields, and different combinations of such kind of choices would yield exponentially many patterns. We need an efficient method for filtering out most of the low-quality patterns without evaluating them.
	\item \emph{Structure Semantics.} 
	Structure extraction is not simply about identifying patterns that can partition the identified records 
	into formatting components (or delimiters), and various pieces of information to be extracted, 
	as the ultimate goal is to transform the log datasets into structured relational datasets. 
	Finding an appropriate structure for this purpose (i.e., making sure that resulting structured datasets are interpretable to users) 
	requires not only a good scoring metric, but also well-designed structure refinement techniques. 
\end{denselist}

\noindent Overall, \algname can automatically extract structure from log datasets without any human supervision. Compared to unsupervised adaptations of semi-supervised structure extraction systems~\cite{recordbreaker, fisher2008dirt}, \algname makes fewer assumptions regarding the structure of the dataset, and therefore is much more applicable towards extracting from log datasets: as shown in our experimental evaluation,  \algname can successfully extract structure from all of the datasets used in Fisher et al.'s work~\cite{fisher2008dirt}, and can achieve $95$\% extraction accuracy on automatically collected log datasets from GitHub, while RecordBreaker~\cite{recordbreaker} can only achieve $29$\% extraction accuracy on the same dataset collection---{\em a substantial 66\% increase.} 
\algname is also efficient and scales well to large datasets: the average running time for small datasets ($<50$MB) is less than $20$ seconds; even for datasets of size more than $100$MB, \algname can still complete extraction within a few minutes. 
The main time spent by \algname for large datasets 
is in extraction (which is eminently parallelizable), and identifying an appropriate structure can be done much faster. 
\new{Via a user study, we demonstrate that \algname is able to generate near-perfect extraction results, compared
to the output of RecordBreaker or supervised extraction on the raw dataset, especially when dealing with real log record datasets with noise.
Our study indicates that \algname can be a useful starting point for supervised extraction
as well, beyond the applicability to large data lakes.}

\vspace{5pt}
\noindent \textbf{Paper Outline.}\enspace The rest of this paper is organized as follows:
\begin{denselist}
	\item In Section~\ref{sec_problem_definition}, we formally define the problem of unsupervised structure extraction.
	\item In Section~\ref{sec_assumption}, we identify key assumptions that will help us solve the problem in a tractable manner. We also compare the assumptions made in our work with those in prior works to demonstrate why \algname is better tailored towards structure extraction from log datasets.
	\item In Section~\ref{sec_alg}, we present \algname, our structure extraction algorithm, and analyze its time complexity and correctness. 
	\item In Section~\ref{sec_experiment}, we experimentally evaluate \algname on $25$ typical datasets and automatically collected log datasets from GitHub, and demonstrate the efficiency, effectiveness, and robustness of \algname for log dataset structure extraction.
	\item \new{In Section~\ref{sec_user_study}, we conduct a user study to compare the extraction results of \algname with RecordBreaker. 
	We show that \algname can handle different types of datasets well, while RecordBreaker requires substantial user supervision for most multi-line record datasets, especially when there is noise present.}
\end{denselist}

\section{Problem Definition}\label{sec_problem_definition}

We now formally define the problem of (unsupervised) structure extraction from log datasets
and introduce related concepts, starting with the concepts of {\em record templates} 
and {\em instantiated records}:

\begin{definition}[Record Template/Instantiated Record]\label{def_record_template}
A {\em record template} is a string that 
contains one or more instances 
of the field placeholder character---a special type of character, denoted as `F' in this paper---along with other characters. 
An {\em instantiated record} is a string with 
no field placeholder character. 
We say an instantiated record {\em can be generated} from 
a record template iff it can be constructed by replacing field placeholder 
characters in the record template with strings containing no field placeholder characters.
\end{definition}

\noindent Given an instantiated record and a record template, 
we can now define the 
concept of {\em field values} as follows:

\begin{definition}[Field Values]
For any pair of instantiated record $R$ and record template $RT$, 
if $R$ can be generated from $RT$, then the replacement strings in $R$ 
for the field placeholder characters are called the {\em field values} 
of $R$ for $RT$. 
When the context is clear, we simply call them the field values of $R$ 
or just the field values.
\end{definition}

\begin{figure}[h]
\vspace{-10pt}
\begin{center}
\includegraphics[width = 4.5cm]{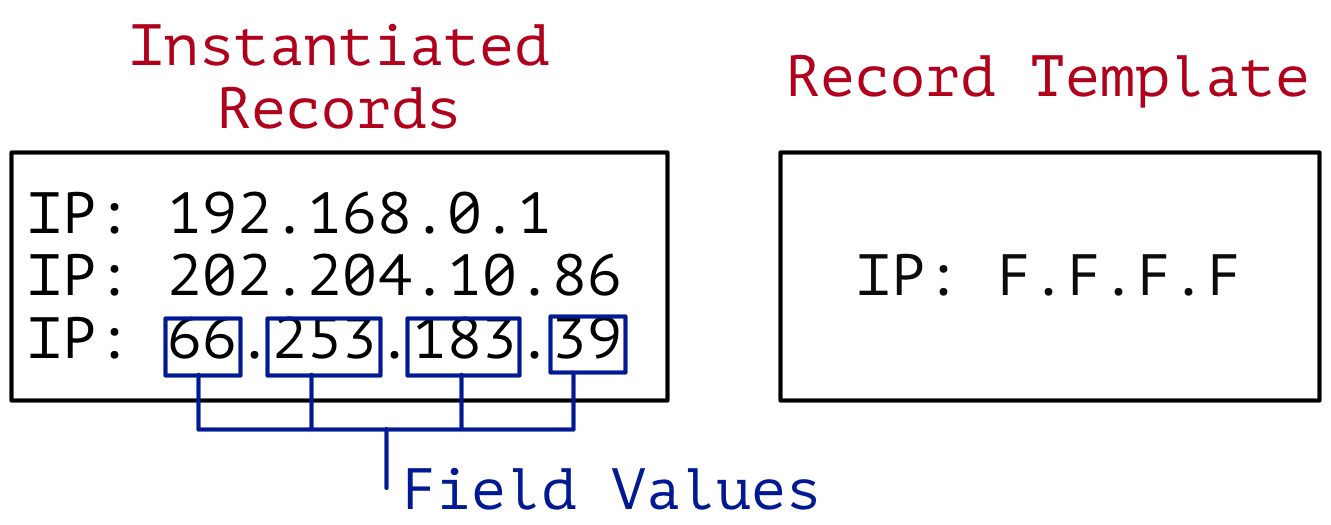}
\end{center}
\vspace{-15pt}
\caption{Record Template Illustration}\label{fig_record_template_definition}
\vspace{-10pt}
\end{figure}

\begin{figure}[h]
	\begin{center}
		\includegraphics[width = 5cm]{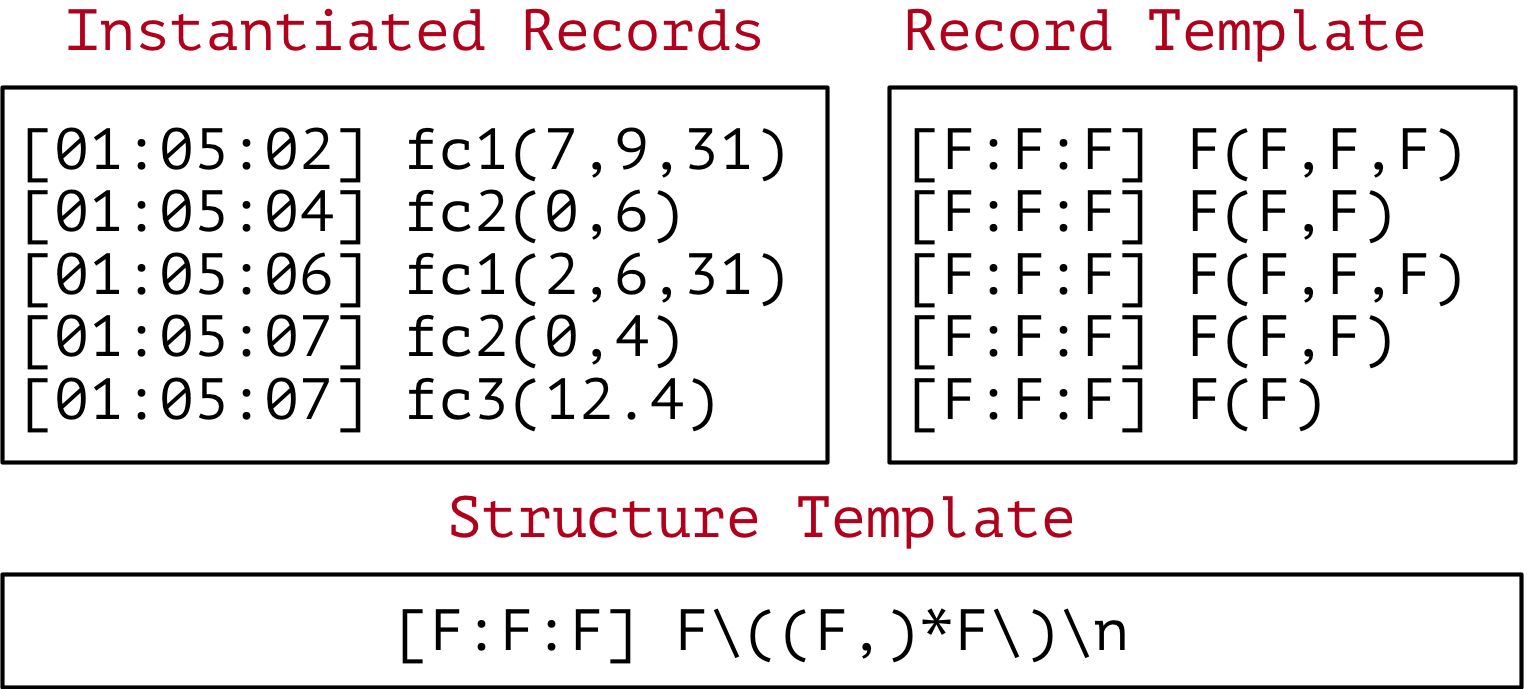}
	\end{center}
	\vspace{-10pt}
	\caption{Structural Uncertainty of Record Templates}\label{fig_record_template_variation}
	\techreport{\vspace{-5pt}}
	\paper{\vspace{-15pt}}
\end{figure}

These definitions are illustrated via an example 
in Figure~\ref{fig_record_template_definition}. 
As we can see, the instantiated records on the left hand side are generated 
by replacing the placeholder character \verb|`F'| in the record template on 
the right hand side with concrete values. 
The data items replacing the placeholder character 
(e.g., $192,168,0,1,\ldots$) are the field values to be extracted from the dataset.

There are often many record templates that could correspond to a given dataset.
Figure~\ref{fig_record_template_variation} 
illustrates an example wherein the corresponding record templates 
(i.e., the right hand side) are similar but not exactly the same. 
To characterize this scenario, we define the concept of a {\em structure template}:

\begin{definition}[Structure Template]\label{def_structure_template}
A {\em structure temp\-late} is a regular expression~\cite{sipser2006introduction} 
for record templates. We say the record template $RT$ 
{\em can be generated} from the structure template $ST$ 
iff the regular expression of $ST$ matches the string form of $RT$.
\end{definition}

\noindent Intuitively, a structure template captures minor variations in the structure of records within a dataset via a regular expression.
The bottom of Figure~\ref{fig_record_template_variation} shows an example structure template corresponding to the records in the top, capturing minor differences in the record templates such as one, two, or three arguments within parentheses.
Now, we define the concept of a {\em log dataset}:

\begin{definition}[Log Dataset]\label{def_dataset}
A log dataset $\mathcal{D} = \{T, S\}$ consists of two components: the textual component $T$ and the structural component $S$. $S = \{ST_1, ST_2$, $\ldots, ST_k\}$ is a collection of structure templates, and $T$ is a text dataset with the following structure: $T$ can be partitioned into several blocks separated by the end-of-line character `\textbackslash n', and each block is either an instantiated record generated 
from one of the structure templates in $S$, or corresponds to a noise string with no structure.
\end{definition}

\begin{figure}[t]
\vspace{-5pt}
\begin{center}
\includegraphics[width = 6cm]{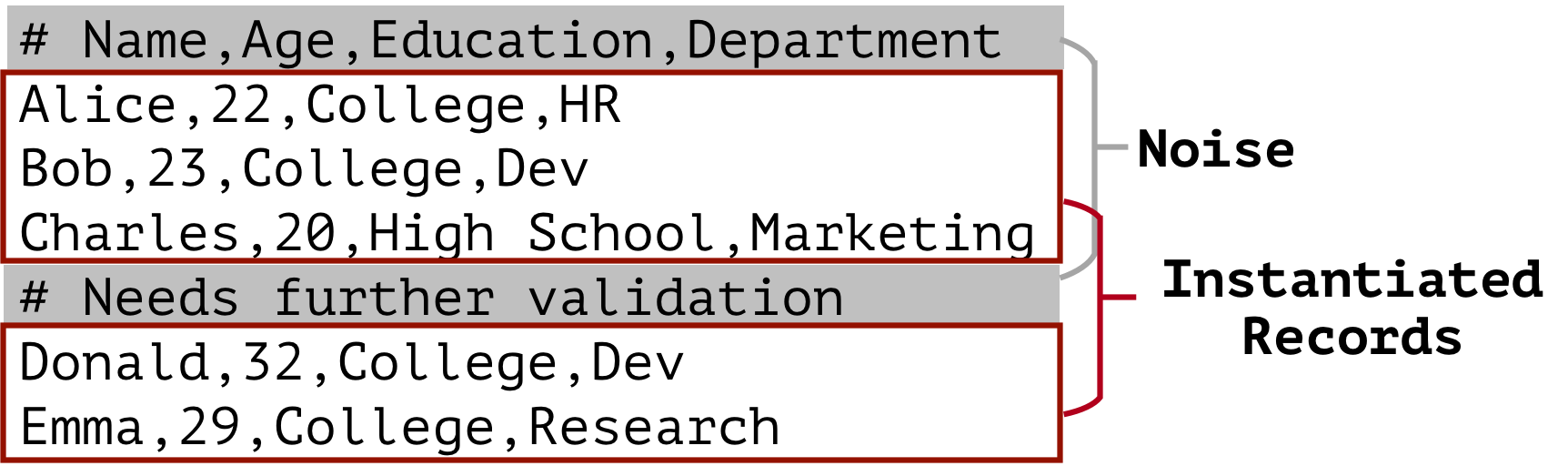}
\end{center}
\vspace{-10pt}
\caption{Log dataset illustration}\label{fig_dataset_definition}
\vspace{-15pt}
\end{figure}

Figure~\ref{fig_dataset_definition} illustrates an example log dataset. 
The parts with a gray background are noise blocks, while the other parts are record blocks.
Noise blocks have no structure within, and are not relevant to the structure extraction problem. 
The requirement that blocks are separated by end-of-line characters in Definition~\ref{def_dataset} is reasonable for log datasets: it seems to be a common practice for programmers to write `\textbackslash n' character at the end of every log line (it holds for every log dataset we have examined). Notice however, that it is not necessary for a record to span just one line, such as the example in Figure~\ref{fig_sl_vs_ml} or Figure~\ref{fig_sample_file}; we only require that the structured components and noise are clearly demarcated.

To formalize the structure extraction problem, we start with an intuitive formulation:

\begin{problem}[Structure Extraction]\label{prob_extract}
For a log dataset $\mathcal{D} = \{T, S\}$ with only $T$ observed but $S$ unknown, recover $S$ and the field values of the instantiated records in $T$. 
\end{problem}


Note that Problem~\ref{prob_extract} is not well-posed: for any given text component $T$, there are infinitely many potential structural components $S$ such that the pair $(T,S)$ obeys Definition~\ref{def_dataset} (for example, the simplest structure template \verb|"F\n"| can pair with any textual component to satisfy Definition~\ref{def_dataset}). Most of these structures are unacceptable from an end-user's point of view. In practice, the structure extraction algorithm needs to discover the most plausible one by designing a scoring system that assigns scores to $(T, S)$ pairs. The scoring system is intended to mimic human judgment: a better score implies that the structure is more plausible from an end-user's point of view. We also adopt this approach in \algname, and the precise {\em \scoreF} $F(T,S)$ we use will be discussed later. Thus, an optimization based formulation of the structure extraction problem is as follows:
\begin{problem}[Structure Extraction (Optimization)]\label{prob_extract_optimization}
For a log dataset $\mathcal{D} = \{T, S\}$ with only $T$ observed but $S$ unknown, find $S$ that optimizes a given \scoreF $F(T, S)$, and extract all the field values of the instantiated records in $T$. 
\end{problem}


\section{Structural Assumptions}\label{sec_assumption}

In Section~\ref{sec_problem_definition}, 
we formalized the structure extraction problem 
as finding the structural component $S$, i.e., 
a collection of regular expressions, 
that best explains or generates the textual component $T$, i.e., 
the one that achieves the highest regularity score $F(T,S)$. 
However, in practice, it is computationally infeasible 
to search over the entire 
space of all possible regular expressions. 
Therefore, it is necessary for structure extraction 
systems---even semi-supervised ones---to 
make additional assumptions on the structural component~\cite{fisher2008dirt,le2014flashextract}. 
These assumptions restrict the search space of 
potential structure templates, 
thereby serving the following two purposes:
\begin{denselist}
	\item To enforce human intuition upon the searching procedure. 
	Structure templates following such assumptions are more likely to be the acceptable 
	from an end-user's point of view. In particular, log files have a regular repeating structure, since they were generated by a computer program and meant to contain all relevant information to be extracted via a computer program or script. Our assumptions serve to codify these principles. 
	\item To reduce the complexity of search space of 
	the structural component, 
	making the structure extraction problem more tractable.
\end{denselist}
In \algname, we make three assumptions 
regarding the structure of the dataset, described next.
The validity of these assumptions will be verified in Section~\ref{sec_github_exp}. We will also compare
these assumptions with the ones made by RecordBreaker~\cite{recordbreaker} at the end of this section.

\vspace{-3pt}

\subsection{Coverage Threshold Assumption}
Here is the first assumption, which is very intuitive:
\begin{assumption}[Coverage Threshold]\label{ass_coverage}
	The coverage of every structure template $ST_i \in S$ 
	should be at least $\alpha\%$. The coverage of structure template 
	$ST$ is defined as the total length (i.e., total number of characters) 
	of the instantiated records of $ST$.
\end{assumption}

\noindent \textbf{Explanation.} Assumption~\ref{ass_coverage} states that log datasets don't typically contain a large number of different structure templates within, 
and thereby each structure template should cover a significant 
portion of the dataset. Note that a structure template is itself a regular expression that can capture a multitude of record templates, so this is not a severe restriction.
The coverage threshold assumption allows us to prune 
out most unreasonable structure template candidates. We will discuss the impact of varying the parameter $\alpha$ in our experiments.


\subsection{Non-Overlapping Assumption}
The second assumption we make is the following:
\vspace{-3pt}
\begin{assumption}[Non-Overlapping]\label{ass_main}
	For any structural template $ST$ and any character $c$, one of the following is true:
	\begin{itemize}
		\item for any record template $RT$ generated from $ST$, $c \notin RT$.
		\item for any instantiated record $R$ generated from $ST$, no field values of $R$ contains $c$. 
	\end{itemize}
\end{assumption}
\vspace{-3pt}

\noindent \textbf{Explanation.} 
Assumption~\ref{ass_main} states that the formatting characters of records cannot be mixed with field values. Intuitively, this makes sense because in practice these records are manually extracted via scripts, and these scripts use delimiters to extract field values. 

To formally explain this, we first define some notation: we let {\em RT-CharSet} denote the set of characters in record templates, while {\em F-CharSet} denotes the set of characters in field values.
Under this notation, Assumption~\ref{ass_main} can be simply stated as: For any structure template $ST$, there exists two disjoint character sets $A(ST)$ and $B(ST)$, such that for any instantiated record $R$ of $ST$, we have {\em RT-CharSet}$(R) \subseteq A(ST)$ and {\em F-CharSet}$(R) \subseteq B(ST)$.
In this paper, we further assume that {\em RT-CharSet} contains only special characters. In other words, we predefine a collection of special 
characters {\em RT-CharSet-Candidate}, and assume that {\em RT-CharSet}$(R)$ $\subseteq$ {\em RT-CharSet-Candidate} for all records $R$.

Assumption~\ref{ass_main} plays an important role: 
it allows us to extract the record template directly 
from an instantiated record 
given the corresponding character set of the record templates, 
and efficiently extract matches for a given
structure template from the dataset.

\vspace{5pt}
\noindent \textbf{Justification of Assumption.}
Assumption~\ref{ass_main} is a relatively strong assumption. To compensate for this, the structural form assumption in Section~\ref{sec_ass_form} (discussed next) is sufficiently flexible such that even for many datasets that seemingly violate this assumption, we can still get reasonable results.
 
For example, consider the record template \verb|F,"F",F|. If the field value surrounded by the quotes contains the comma character, then Assumption~\ref{ass_main} would be violated.
However, \algname will still be able to recognize several different record templates in the following, depending on the number of commas in the middle field value:

\begin{verbatim}
      F,"F",F       F,"F,F",F      F,"F,F,F",F
\end{verbatim}

Since all the above record templates can be generated from the same 
structure template \verb|F,"(F,)*F",F|, they will still be recognized as the same record type. We formalize the space of structure templates next.

%

\begin{figure}[t]
	\techreport{\vspace{-10pt}}
	\paper{\vspace{-25pt}}
	\begin{center}
		\includegraphics[width = 5cm]{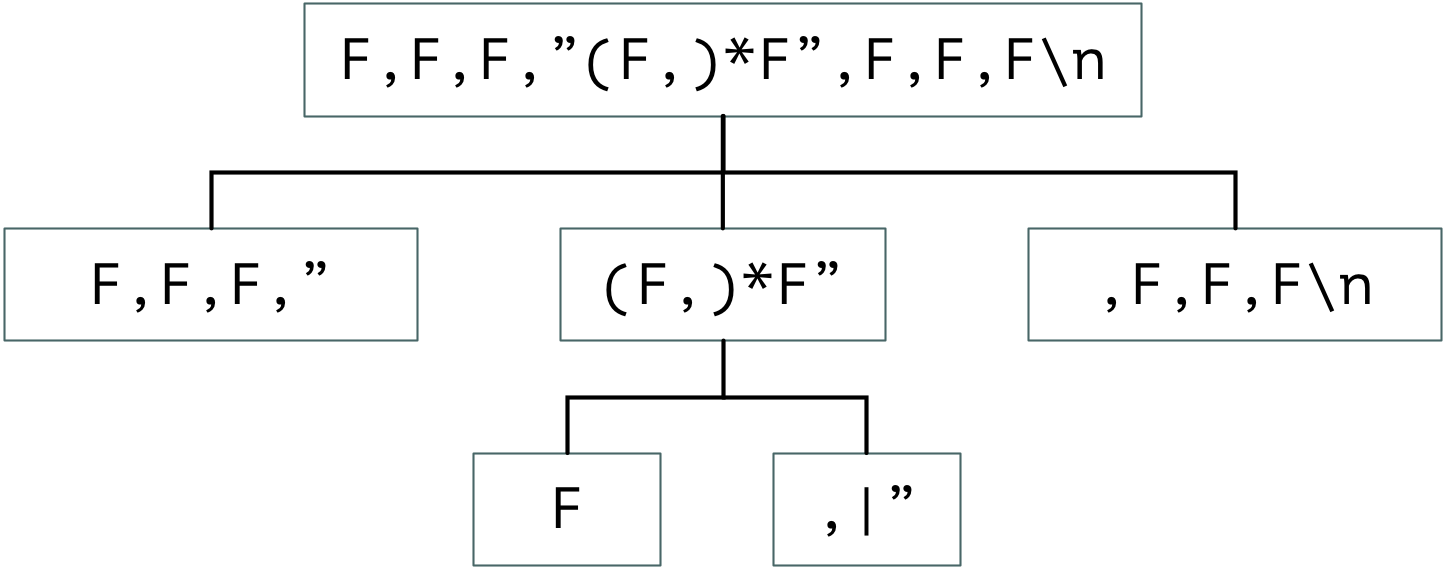}
	\end{center}
	\vspace{-10pt}
	\caption{Structural Form Assumption}\label{fig_structure_template_definition}
	\techreport{\vspace{-10pt}}
	\paper{\vspace{-15pt}}
\end{figure}

\subsection{Structural Form Assumption}\label{sec_ass_form}
The following assumption restricts the forms of structure templates:
\vspace{-10pt}
\begin{assumption}[Structural Form]\label{ass_form}
	Every structure template is a regular expression that has one of the following forms:
	\begin{enumerate}
	\itemsep0.1em
		\item Array: \verb|({regexA}x)*{regexA}y| \\
		where \verb|{regexA}| is another regular expression satisfying Assumption~\ref{ass_form}, and \verb|x| and \verb|y| are different characters. 
		\item Struct: \verb|{regexA}{regexB}{regexC}....| \\
		where \verb|{regexA}{regexB}{regexC}....| is a sequence of regular expressions, and each of them is either a simple string or another regular expression satisfying Assumption~\ref{ass_form}.
	\end{enumerate}
\end{assumption}

\noindent \textbf{Explanation.} Assumption~\ref{ass_form} states that records in log datasets are laid out from left to right, with nesting. Formally, the Array-type regular expression is intended to characterize lists of objects. 
For example, the structure template \verb|[F,F,F,...,F]| can be represented by a prefix \verb|[| and an array-type regular expression \verb|(F,)*F]|.
Thus, Assumption~\ref{ass_form} essentially states that each structure template 
must follow a special tree-style structure. An example tree structure for the structure template \verb|F,F,F,"(F,)*F",F,F,F\n| is illustrated in Figure~\ref{fig_structure_template_definition}. As we can see, 
the root node in this tree is a Struct node, with three children nodes (level 2 in Figure~\ref{fig_structure_template_definition}). The second node in level $2$ in Figure~\ref{fig_structure_template_definition} is an Array regular expression node that has two children nodes (level 3 in Figure~\ref{fig_structure_template_definition}): 
the left child is the regular expression part, and right child is the terminating character part.

\begin{figure}[t]
	\begin{center}
		\includegraphics[width = 6cm]{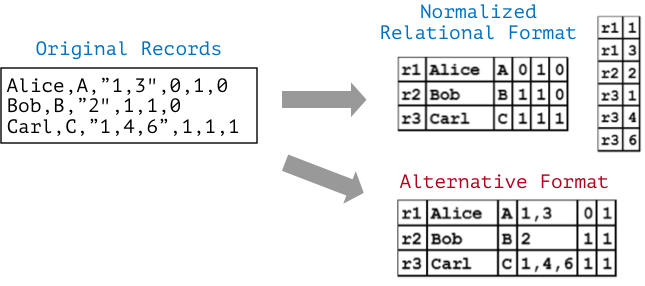}
	\end{center}
	\vspace{-10pt}
	\caption{Extracted Relational Dataset}\label{fig_relational_format}
	\vspace{-10pt}
\end{figure}

\new{We can store all of the extracted records in a relational format based 
on the tree-structure in Assumption~\ref{ass_form}. 
Figure~\ref{fig_relational_format} demonstrates this procedure: 
the instantiated records on the left hand side 
are generated from the structure template in Figure~\ref{fig_structure_template_definition}, 
and the right hand side depicts two representations for the relational dataset: one, a normalized relational format, 
and the other, a denormalized format that uses arrays. 
As we can see, for the normalized format, each field-placeholder character `F' 
in the structure template corresponds to one column in the relational dataset, 
and the correspondence between non-leaf nodes and their parents are captured using foreign-key references. 
\algname can generate either representation, both of which contain 
all of the extracted information, and can be utilized by downstream applications.
}


\new{Our template language in Assumption~\ref{ass_form} is basically the same as in LearnPADS~\cite{fisher2008dirt},
	except that we do not use a union type-constructor. However, compared to LearnPADS, our definition of a log dataset in 
	Definition~\ref{def_dataset} and the corresponding problem formulation is novel: in Definition~\ref{def_dataset},
	we defined a log dataset as concatenation of instantiations of multiple types of structure templates plus potentially
	heavy noise. In contrast, LearnPADS assumes the log dataset to be a well-defined list of chunked records. 
	As described earlier, a key 
	difference is that we no longer assume that record boundaries are known beforehand. This
	difference leads to a very different algorithmic solution as we shall see next.
}

\techreport{
\vspace{5pt}
\noindent \textbf{Remark.} 
The regular expression form depicted in Assumption~\ref{ass_form} can be rewritten as an equivalent LL(1) grammar~\cite{grune2007parsing}. Therefore, after finding the optimal structure template using \algname, the actual extraction procedure can be done by the canonical LL(1) parser in linear time.
}

\subsection{Assumption Comparison}

Here we compare the assumptions made in \algname with those in RecordBreaker~\cite{recordbreaker}. The structural form assumption (Assumption~\ref{ass_form}) has an equivalent counterpart in RecordBreaker. RecordBreaker also makes a stronger version of Assumption~\ref{ass_main}, together with another additional assumption regarding record boundaries:
\vspace{-5pt}
\begin{assumption}[Boundary]\label{ass_boundary}
	The boundaries of records can be easily identified beforehand.
\end{assumption}
\vspace{-5pt}
\begin{assumption}[Tokenization]\label{ass_tokenize}
	Each record can be tokenized beforehand, such that each token 
	is either part of a field-value, or part of the structure template. 
	In other words, in addition to Assumption~\ref{ass_main}, it is further assumed that {\em RT-CharSet} for all records are predetermined in advance\footnote{\scriptsize The only difference between Fisher's algorithm~\cite{fisher2008dirt} and RecordBreaker~\cite{recordbreaker}  is the treatment of this assumption: Fisher et al. assume that {\em RT-CharSet-Candidate} is given by the user for each dataset; RecordBreaker compiled a predetermined character set, making their program unsupervised.}:
	$\forall R, \textit{RT-CharSet}(R) = \textit{RT-CharSet-Candidate}. $
\end{assumption}
\vspace{-5pt}

\begin{table}[h]
\centering
\scriptsize
\vspace{-10pt}
\begin{tabular}{|c|c|c|}
\hline
Assumption & RecordBreaker & \algname \\
\hline
Coverage Threshold & No & Yes \\
Non-overlapping & Yes & Yes \\
Structural Form & Yes & Yes \\
Boundary & Yes & No \\
Tokenization & Yes & No \\
\hline
\end{tabular}
\caption{The Assumption Comparison Chart}\label{tbl_assumption}
\vspace{-20pt}
\end{table}

Table~\ref{tbl_assumption} compares the assumptions in RecordBreaker and \algname. As discussed in the introduction, the two additional assumptions in RecordBreaker are rather restrictive for log datasets. This is further verified in our experiments:
about 31\% of the log datasets we automatically collected from GitHub (details in Section~\ref{sec_github_exp}) do not satisfy these assumptions. In comparison, the additional assumption made in \algname is much milder: due to the coverage threshold assumption, we will only extract from ``popular'' structure templates rather than all of them. In most practical settings, such a restriction wouldn't cause any problems.

\section{The {\large \algname} Algorithm}\label{sec_alg}

In Section~\ref{sec_problem_definition}, we defined the structure extraction problem as the problem of finding the structural component $S$ that optimizes a given \scoreF $F(T, S)$ given the observed textual 
component $T$. \new{Recall that $T$ has the following form (Definition~\ref{def_dataset}):
$$T = B_1\backslash nB_2\backslash n \ldots \backslash n B_n$$
where each block $B_i$ is either a noise block or an instantiated record generated from one of the structure templates in $S$. Due to the extremely large search space of structure templates described in Assumption~\ref{ass_form}, exhaustive search is not an option, and it is necessary to use the information within $T$ while searching for potential structure templates. }
	
\new{Most prior unsupervised structure extraction algorithms~\cite{fisher2008dirt, recordbreaker, elmeleegy2009harvesting} assume that the record boundaries are known beforehand. These algorithms are usually based on the idea of summarization: given all the examples generated from the structure template, the algorithm tries to find the structure template by seeking out the common patterns among records. However, as mentioned previously, the record boundaries within log datasets are usually unknown. which makes these algorithms not directly applicable to log datasets. Furthermore, the task of finding record boundaries is itself also not easy: without knowing the record characteristics first, it is very difficult to pinpoint the exact location of record boundaries, especially with the presence of heavy noise.}

\begin{figure}[t]
	\vspace{-20pt}
	\begin{center}
		\includegraphics[width = 8cm]{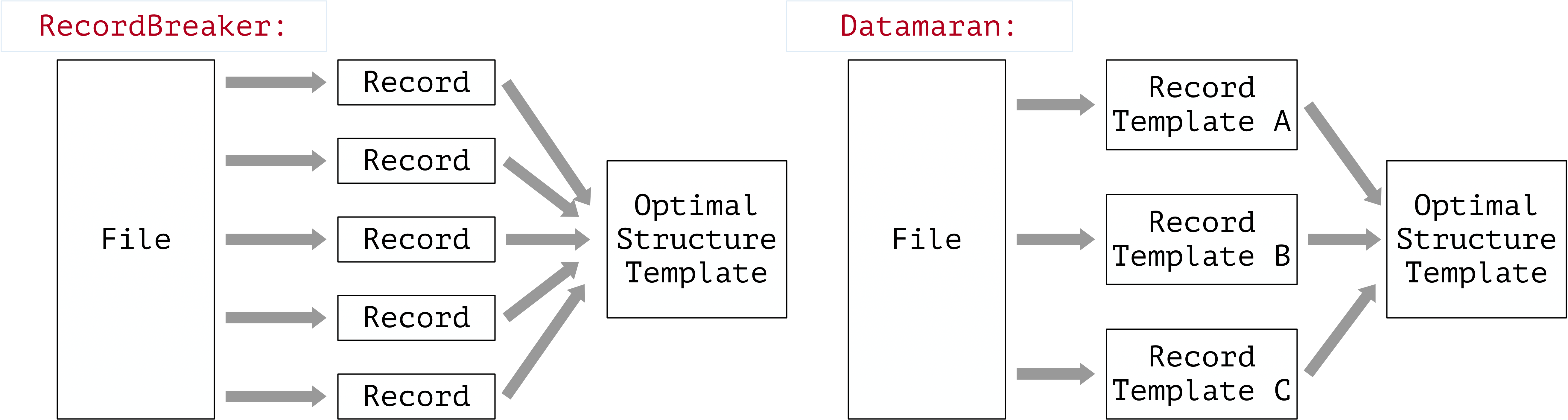}
	\end{center}
	\vspace{-10pt}
	\caption{\algname vs. RecordBreaker}\label{fig_approach_flow}
	\vspace{-10pt}
\end{figure}

\begin{figure*}[!t]
\vspace{-20pt}
	\begin{center}
		\includegraphics[width = 12cm]{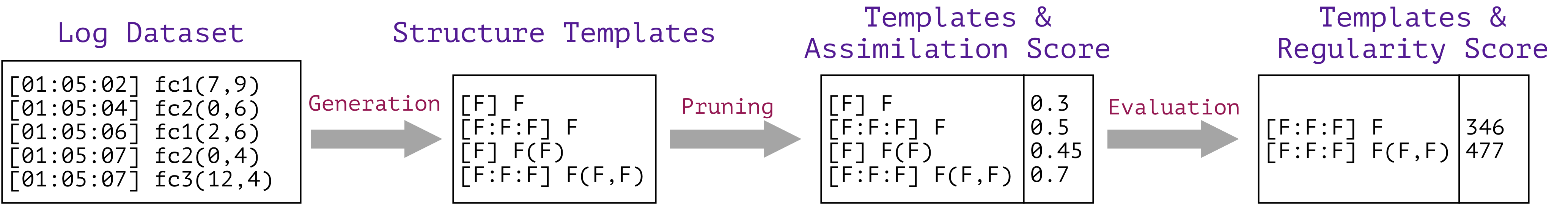}
	\end{center}
	\vspace{-10pt}
	\caption{The Workflow of \algname}\label{fig_workflow}
	\vspace{-10pt}
\end{figure*}

Given the difficulty associated with identifying record boundaries, a different approach is used by \algname: \new{\algname first generates a large collection of structure template candidates directly from the dataset (without actually identifying the record boundaries), and then evaluates the most promising ones to find the optimal structure template.} Figure~\ref{fig_approach_flow} illustrates the conceptual differences between \algname and prior approaches such as RecordBreaker~\cite{recordbreaker}. Concretely, \algname algorithm consists of the following three steps, as illustrated in Figure~\ref{fig_workflow}:

\vspace{-3pt}
\begin{itemize}
	\itemsep0.1em
	\item \emph{Generation.} The first step is to search for candidate structure templates that satisfy the coverage threshold assumption (Assumption~\ref{ass_coverage}). \new{To achieve this, we first extract a large collection of structure templates from potential records (i.e., consecutive lines in the dataset), then insert these structure templates into a hash-table to find repeated ones.}
	\item \emph{Pruning.} The second step is to prune out most of the candidates found in the previous step, such that \new{we only need to evaluate the \score of a small number of candidates}. To achieve this, we designed an {\em \apprxscoreF} $G$, a built-in \scoreF that can be evaluated very efficiently. \new{Intuitively, this \apprxscoreF tries to filter out all of the redundant structure templates derived by removing some structural details from the true structure templates.} We then retain the candidates with highest \apprxscore $G(T,S)$ for the final evaluation.
	\item \emph{Evaluation.} During the final step, we apply two structure refinement techniques to the remaining structure templates after the pruning step, and then evaluate their \score to find the one with the highest $F(T,S)$.
\end{itemize} 
\vspace{-3pt}
\new{The primary algorithmic contributions of \algname are the implementations of {\em generation} and {\em pruning} step:  (a) for the {\em generation} step, extracting structure templates directly from potential records is highly nontrivial due to the possible variations of field values and record template structures (see Assumption~\ref{ass_form}), and Assumption~\ref{ass_main} plays an important role in this step; (b) for the {\em pruning} step, the {\em \apprxscoreF} requires careful design: it has to be simple enough so that we can evaluate it efficiently, while being effective enough to be able to prune out most of the low-quality redundant candidates. Our final design is based on several iterations, and is not straightforward at first glance. }

The details of the \algname algorithm will be discussed in the rest of this section: in Section~\ref{sec_generation}, we describe the algorithm for efficiently finding structure templates satisfying the coverage threshold assumption (Assumption~\ref{ass_coverage}); in Section~\ref{sec_pruning}, we describe our \apprxscoreF and discuss the intuition behind its design; in Section~\ref{sec_refine}, we describe two structure refinement techniques that are applied during the evaluation step; in Section~\ref{sec_theory}, we analyze the time complexity of \algname and characterize the conditions under which the correctness of \algname can be guaranteed. There are some additional algorithmic details of \algname that will not be discussed in this section due to page limitations, and they can be found in the appendix.

\vspace{3pt}

\noindent \textbf{The Regularity Scoring Function.} In \algname, we assume the \scoreF $F(T,S)$ is given, and we can access it through a function call. The design of \algname is independent of the choice of this scoring function: we can plug in any reasonable scoring function into \algname, and the algorithm would function as before. In this sense, the primary contribution of \algname is an efficient and scalable method to optimize any reasonable scoring function.

However, for completeness, we will present the details of the minimum description length~\cite{barron1998minimum} \scoreF that we use in our implementation in the appendix, and we demonstrate that it does well empirically in Section~\ref{sec_experiment}. That said, through the rest of this section, we assume this function is given and it mimics human judgment regarding the quality of structure templates.

\vspace{3pt}
\noindent  \textbf{Notation.} Table~\ref{tbl_notation} lists the notations used in \algname. The first $3$ symbols are parameters in \algname, while the last $5$ symbols represent dataset-dependent values. We 
will describe each of these parameters later on.

\begin{table}[h]
\vspace{-5pt}
\small
\centering
	\begin{tabular}{c|l}
		Symbol & Description \\
		\hline
		\hline
		$M$ & The number of structure templates retained after pruning \\
		\hline
		$L$ & The maximum span of records (i.e., the maximum\\
		& number of lines each record can span) \\
		\hline
		$\alpha$ & The minimum coverage threshold for records \\
		\hline
		$n$ & The total number of lines in the dataset \\
		\hline
		$K$ & The number of structure templates retained after generation \\
		\hline
		$T_{data}$ & The total size of the dataset \\
		\hline
		$S_{data}$ & The amount of data sampled during all three steps\\
		\hline
		$c$ & The number of special characters (i.e., characters \\
		& in {\em RT-CharSet-Candidate}) appearing in the dataset \\
	\end{tabular}
	\caption{Notation Summary}\label{tbl_notation}
	\vspace{-20pt}
\end{table}

\subsection{The Generation Step}\label{sec_generation}

\begin{figure*}[t]
	\vspace{-20pt}
	\begin{center}
		\includegraphics[width = 12cm]{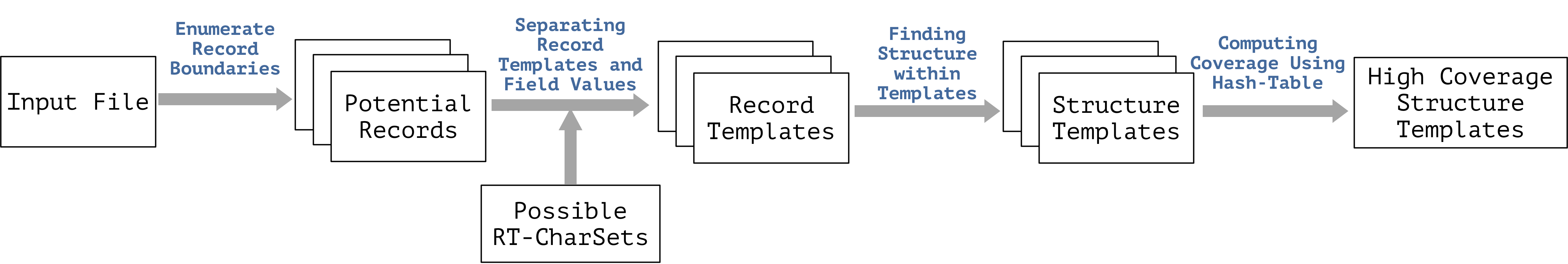}
	\end{center}
	\vspace{-10pt}
	\caption{The Generation Step Workflow}\label{fig_generation_workflow}
	\vspace{-10pt}
\end{figure*}

In the generation step, we find structure templates satisfying Assumption~\ref{ass_coverage} (i.e., those with at least $\alpha$\% coverage). At a high level, this is achieved by finding repetitive patterns within the dataset. Specifically, \algname uses the following five steps to find structure templates with at least $\alpha\%$ coverage:

\begin{enumerate}
\itemsep0.1em
	\item Enumerate possible values of {\em RT-CharSet} (i.e., the character set in the record templates), and for each such value of {\em RT-CharSet}, run through steps 2-5. 
	\item Enumerate all $O(nL)$ pairs of end-of-line characters \Verb|'\n'| that are close to each other (i.e., at most $L$ lines are between them) in the textual component $T$. 
	For each such pair, treat the content between each pair as an instantiated record, and run steps 3-4. 
	\item Extract the record template from the instantiated record using the value of {\em RT-CharSet}. 
	\item Reduce the record template into a structure template (with the form defined in Assumption~\ref{ass_form}).
	\item Store all of the structure templates generated in step 4 within a hash-table, and then find the ones that satisfy the coverage threshold assumption.
\end{enumerate}
	Figure~\ref{fig_generation_workflow} illustrates the workflow of the generation step. \new{The basic idea behind the generation step is very simple: we first enumerate all possible record boundaries (Step 2), then extract structure templates from the contents between them (Step 3, 4), and finally use a hash-table to find the ones with sufficient coverage (Step 5). Assumption~\ref{ass_main}, which states that {\em RT-CharSet} $\cap$ {\em F-CharSet} $= \emptyset$, is the key assumption that allows us to extract record templates directly from instantiated records (Step 3). Using this assumption, we can separate the field values from formatting characters after enumerating possible values of {\em RT-CharSet} (Step 1). More details of these steps, together with pseudo-code, can be found in Section~\ref{sec_detail} in the appendix. }
	
	\new{The search procedure of generation step is {\em complete}: since we are enumerating all possible record boundaries, all occurrences of the true structure template will be accounted for. Therefore, the hash-bin associated with the true structure template is guaranteed to have sufficient coverage, which ensures that the true structure template will be among the list of candidate structure templates  after the generation step.}
	
	\vspace{5pt}
	\noindent \textbf{Variants of Generation \new{and Sampling Technique}:} There are two different versions of the first sub-step implemented in \algname: the exhaustive version enumerates all possible values while the greedy version searches only a subspace of possible values. Intuitively, these two searching procedures represent a trade-off between accuracy and efficiency: the exhaustive search is slower but gives us better extraction results. 
	\new{Additionally, since the running time of generation scales linearly with respect to the dataset size, it may be very expensive for large datasets. We have used a sampling method to ameliorate this. Details of these two techniques can also be found in Section~\ref{sec_detail} in the appendix. }

\subsection{The Pruning Step}\label{sec_pruning}

Even with the coverage threshold assumption, there are often far too many structure template candidates remaining after the generation step. As a result, it is impossible to evaluate \score $F(T,S)$ for every single one. The purpose of the pruning step is to identify a small promising subset of these candidates to be evaluated in the final evaluation step, and discard the rest. 

In the pruning step, we use {\em \apprxscore} $G(T,S)$ to order the structure templates, so that only the best $M$ ones need to be 
evaluated explicitly in the evaluation step. The \apprxscore estimates the amount of data ``assimilated'' by the structure template (i.e., the amount of data that can be explained by the structure template). Therefore, structure templates with a higher \apprxscore are more likely to also have a higher \score.

\begin{figure}[h]
	\vspace{-10pt}
	\begin{center}
		\includegraphics[width = 7cm]{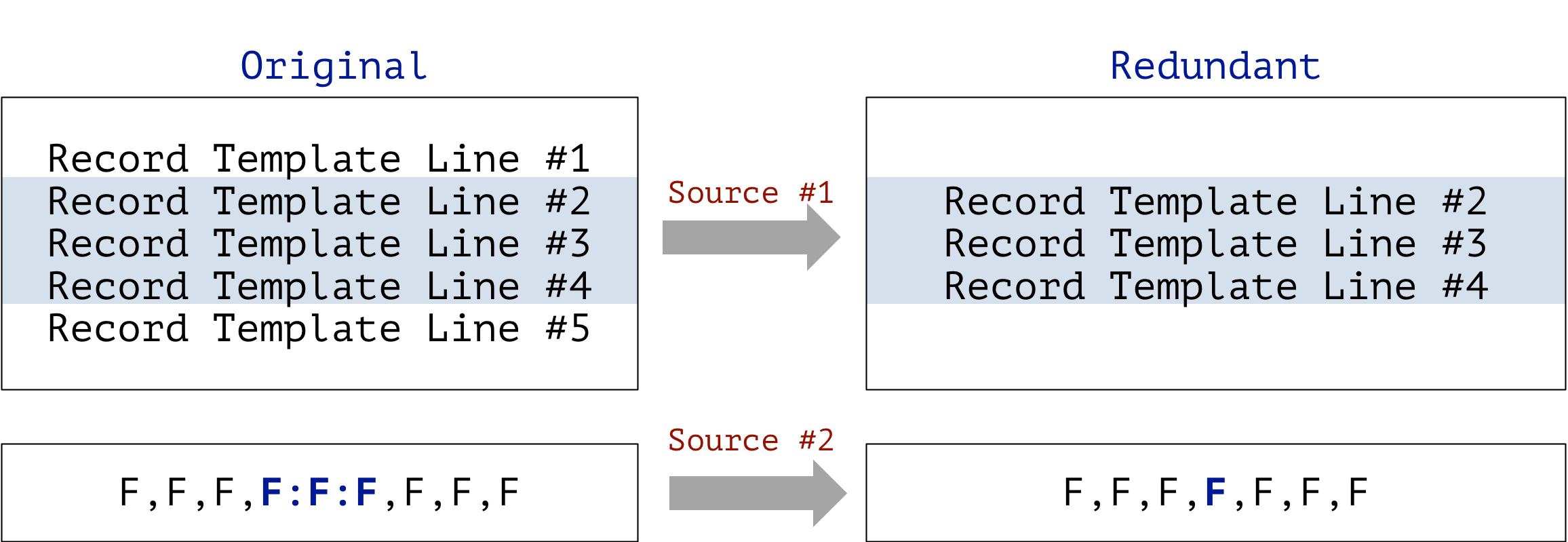}
	\end{center}
	\vspace{-10pt}
	\caption{Two sources of redundancies: (1) subsets of multi-line structure templates; (2) structural parts recognized as field values.}\label{fig_redundancy}
	\vspace{-10pt}
\end{figure}

Before we describe the actual design of our \apprxscoreF, it is helpful to first understand why there are so many structure templates remaining after the generation step. It turns out that most of the redundant structure templates come from two sources as demonstrated in Figure~\ref{fig_redundancy}: (a) when the structure template consists of multiple lines (line 1-5 in Figure~\ref{fig_redundancy} left), any subset of such a structure template would also be captured by the generation step as a legitimate structure template (line 2-4 in Figure~\ref{fig_redundancy} right); (b) when the structure template uses multiple types of characters to separate the field values, simpler structure templates can be recognized if some of those characters are treated as field values as illustrated in Figure~\ref{fig_redundancy} (bottom).

Therefore, a good \apprxscore should be able to distinguish both types of redundancies, and rank the true structure template(s) higher than the redundant ones. At the same time, it should be relatively lightweight to compute. To achieve this, our first component uses the coverage value of structure templates, since we have already computed it during the generation step. However, while the coverage value can effectively distinguish the first source of redundancy, it is not capable of distinguishing the second one. \techreport{As a result, using the coverage value directly as the \apprxscore will not serve our propose.}

To address this shortcoming, we introduce another component 
into the \apprxscore: the {\em Non-Field-Coverage} term, 
which is defined as the total coverage of the structure template 
minus the total coverage of all field values of the structure template 
(i.e., the total length covered by field values in the instantiated records). 
This term computes the total coverage achieved by ``non-field'' characters 
in the template, and can be effectively used to distinguish the second source of redundancy. 
The final \apprxscoreF $G(T,S)$ used in \algname is the following, which filters out all structure templates with either low coverage or low non-field-coverage.
$$G(T,S) = Cov(T,S) \times Non\_Field\_Cov(T,S)$$


\paper{\vspace{-10pt}}
\subsection{Structure Refinement}\label{sec_refine}


To further improve the extraction accuracy, 
we developed two techniques to refine the structure templates. These techniques are applied to all of the top $M$ structure templates during the evaluation step: 
we revise these structure templates, and compare the revised structure templates against the original ones, using the \scoreF, 
replacing them if the score is improved.

\paper{\vspace{-3pt}}
\subsubsection{Array Unfolding}

During the generation step, all of the records are transformed into minimal structure templates, which allowed us to detect repetitive patterns within the dataset. However, there are cases where the minimum structure template is not the optimal structure template.

For instance, in comma-separated values files (*.csv files), all of the records have the form \Verb|"F,F,F,....,F,F\n"| (i.e., a fixed number of field values separated by commas). There are two possible structure templates for these records: the plain struct-type \Verb|"F,F,F,....,F,F\n"| and the array-type \Verb|"(F,)*F\n"|. The plain struct-type template offers a better semantic interpretation in this case (since it implies that the field values are of different types), and also leads to a better \score $F(T,S)$.

\begin{figure}[h]
	\vspace{-10pt}
	\begin{center}
		\includegraphics[width = 4cm]{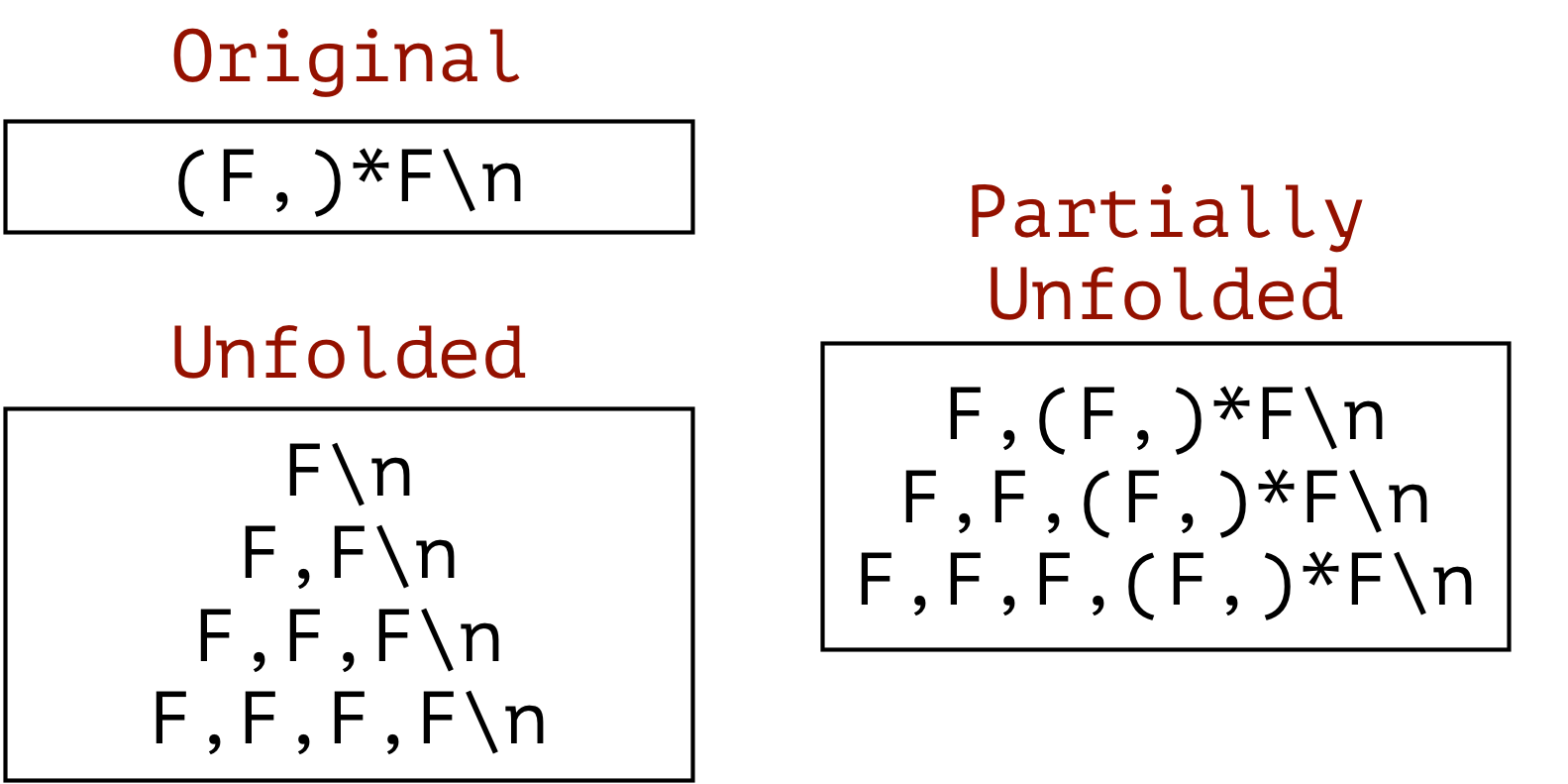}
		\includegraphics[width = 4cm]{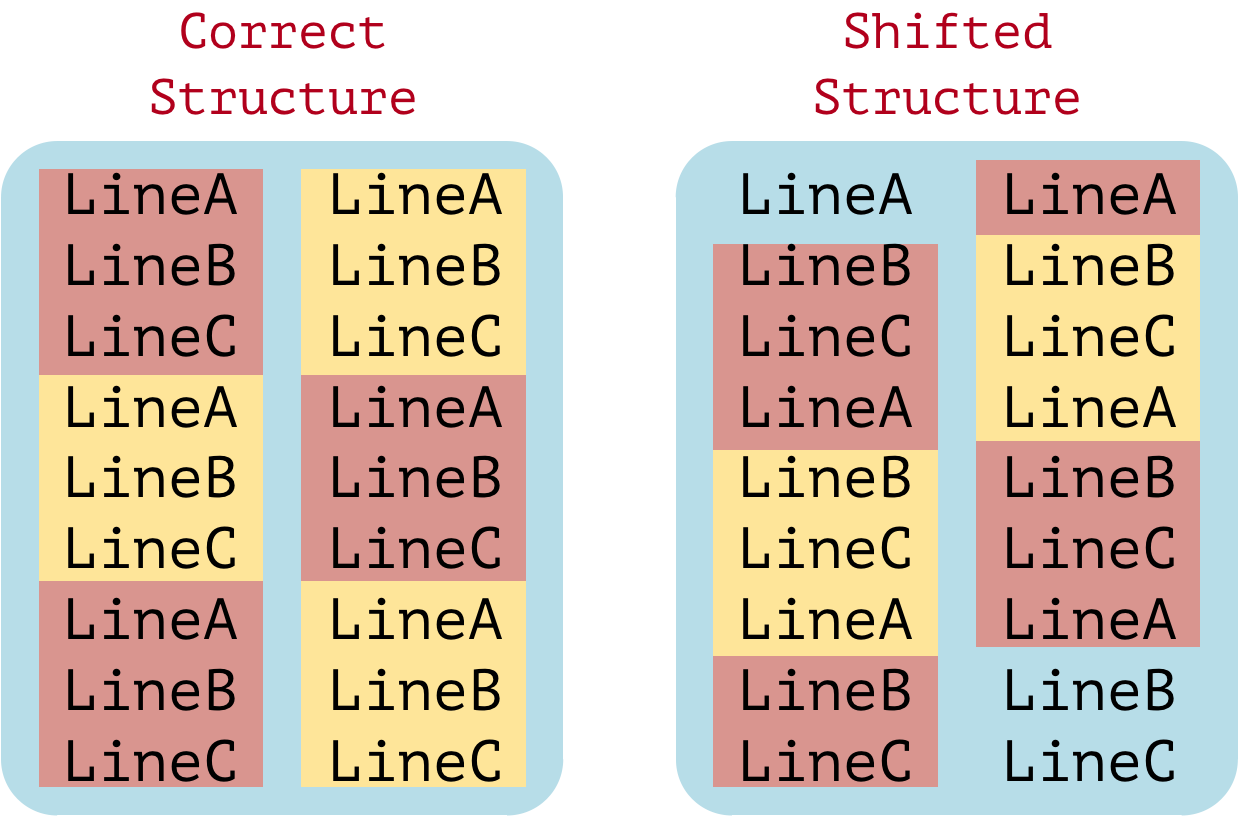}
	\end{center}
	\vspace{-10pt}
	\caption{Array Unfolding (left); Structure Shifting (right)}\label{fig_unfolding_shifting}
	\vspace{-5pt}
\end{figure}

\new{More generally, because of the structure template reduction procedure (step 4 in the generation step), when the optimal structure template is not a minimal structure template, only its reduced form will be found during the generation step. }
To address this, we designed the {\em array unfolding} technique: 
for each array-type regular expression in the structure template, 
we attempt to unfold it by expanding it into a struct-type. Figure~\ref{fig_unfolding_shifting}(a) demonstrates this process: the array-type regular expression at the top of the figure will be unfolded into one of the struct-type regular expression at the bottom of the figure. If any of these unfolded structure templates has a better score than the original, the unfolding would be finalized.

\techreport{
\new{
Partial unfolding, another unfolding mechanism implemented in \algname, is also demonstrated in Figure~\ref{fig_unfolding_shifting}(a). Here, we expand the array-type regular expression while retaining the non-deterministic array-type suffix.}
Partial unfolding is used to handle the cases 
where regular field values are ``mixed in'' with text field values, as in the following example: 

{\small
\noindent \Verb|Apr 24 04:02:24 srv7 snort shutdown succeeded|\\
\Verb|Apr 24 04:02:24 srv7 snort startup succeeded|\\
\Verb|Apr 24 14:44:28 srv7 Disabling nightly yum|
}

\noindent In this example, the first four fields are regular fields, but the last one is a free-text field. The ideal structure template for this example is  \Verb|F F F F| \Verb|(F )*F\textbackslash n|, which can be obtained by applying partial unfolding to the minimum structure template \Verb|(F )*F\textbackslash n|.
}

\paper{\vspace{-3pt}}
\subsubsection{Structure Shifting}

	Typically, the \scoreF $F(S,T)$ evaluates the quality of structure templates using statistics such as coverage value or minimum description length (see Section~\ref{sec_detail}). For most cases, these kinds of score functions can distinguish good structure templates from bad ones. However, there is
	one ambiguity among structure templates that such a \score would fail to detect:
	the cyclic shifting of structure templates.
	Figure~\ref{fig_unfolding_shifting}(b) illustrates this: the \score $F(T,S)$ of the shifted structure template (right hand side in Figure~\ref{fig_unfolding_shifting}(b)) and the score $F(T,S)$ 
	of the correct structure template (left hand side in Figure~\ref{fig_unfolding_shifting}(b)) 
	are usually approximately equal to each other. 


\new{
The structure shifting mechanism in \algname is designed to distinguish such ambiguities: for each structure template, we consider all possible shifted variants, and then find the position of first occurrence for each one of them. We then pick the one with the earliest first occurrence, which intuitively is most likely the correct structure.
}

\subsection{Theoretical Analysis}\label{sec_theory}


\subsubsection{Time Complexity}\label{sec_time_complexity}

Table~\ref{tbl_time_complexity} lists the time complexity of the three 
steps in \algname respectively\footnote{\scriptsize There are two variants of the search procedure for enumerating RT-CharSet in the generation step, see Section~\ref{sec_detail} for details.}. An explanation for the symbols can be found in Table~\ref{tbl_notation}. Note that for large datasets, we would utilize sampling for both the generation and evaluation step (details in Section~\ref{sec_detail}), and therefore $S_{data}$ is upper-bounded by a large constant.
In such cases, the running time of our algorithm is dominated by the actual data extraction procedure.

\begin{table}[h]
	\centering
	\small
	\vspace{-5pt}
	\begin{tabular}{c|c}
		Step & Time Complexity \\
		\hline
		Generation Step & $O(S_{data}L2^c)$ or $O(S_{data}Lc^2)$  \\
		Pruning Step & $O(K \log K)$ \\
		Evaluation Step & $O(MS_{data})$ \\
		Data Extraction & $O(T_{data})$ \\
	\end{tabular}
	\caption{Time Complexity of the Three Steps in \algname}\label{tbl_time_complexity}
	\vspace{-20pt}
\end{table}


\subsubsection{Correctness Guarantee}\label{sec_correctness}

\algname is designed to tolerate noise blocks and variations within record structures and field values. Here we characterize three conditions that are sufficient for guaranteeing the correctness of \algname: 

\begin{theorem}\label{thm_performance}
	For a log dataset $D = \{T, S\}$ with only $T$ observed, 
	if the following conditions are all met:
	\begin{enumerate}[(a)]
		\itemsep 0.1em
		\item One of the structure templates in $S$ (denote it as $ST_0$) has the highest coverage and non-field-coverage (defined in Section~\ref{sec_pruning}) among all structure templates.
		\item For at least $\alpha\%$ of the instantiated records, the minimum structure template for them is $ST_0$.
		\item $ST_0$ has the best \score among all structure templates.
	\end{enumerate}
	
	Then \algname is guaranteed to return $ST_0$ as the optimal structure template.
\end{theorem}

The proof can be found in Section~\ref{sec_thm_prf} in the appendix. For most practical settings, condition (b) is automatically met. Condition (c) requires a carefully designed score function, which is not the focus of this paper. As for condition (a), intuitively it requires the structure templates in $S$ to be sufficiently different from each other, and the field values and noise blocks are sufficiently random. If all of these conditions are satisfied, then Theorem~\ref{thm_performance} would guarantee the correctness of \algname.

\section{Performance Evaluation}\label{sec_experiment}

In this section, we experimentally evaluate the performance of \algname. 
The experiments are conducted on two sets of datasets serving different purposes:
\vspace{3pt}
\begin{denselist}
	\item \textbf{Manually collected log datasets (Section~\ref{sec_example_dataset}).} We collected $25$ datasets, including the entire set of $15$ datasets used by Fisher et al.~\cite{fisher2008dirt} and $10$ others from various sources (details in Section~\ref{sec_example_dataset}).
	These datasets cover a wide variety of structural formats 
	and possess different characteristics (e.g., file size or structural complexity). 
	We use these datasets to study various properties of \algname 
	such as effectiveness, efficiency, parameter sensitivity, and scalability. 
	\item \textbf{GitHub log datasets (Section~\ref{sec_github_exp}).} We crawled a collection of $100$ log datasets automatically 
	from public GitHub repositories. 
	These datasets reflect the properties of real-world data lakes. 
	We use these datasets to study the properties of data lakes ``in the wild'',
	as well as the utility of \algname in such settings. This collection of datasets can be viewed as a benchmark for further research.
\end{denselist}
\vspace{3pt}

 \noindent \textbf{\algname Settings:} \algname is implemented in C++\techreport{ and compiled under Visual Studio 2015}.
The default values for the three parameters in \algname are: $\alpha = 10\%$ 
(the coverage threshold parameter); $L = 10$ (the upper bound of record span); 
$M = 50$ (the number of remaining structure templates after the pruning step). 
These default values are used in all of our experiments except for our parameter sensitivity experiments.
\vspace{3pt}

\noindent \textbf{RecordBreaker~\cite{recordbreaker} Settings:} Despite our best attempts, we were unable to install or run the open-source version of RecordBreaker~\cite{recordbreaker}. Therefore, we decided to faithfully reimplement RecordBreaker in C++ for our comparison. At the first step, RecordBreaker relies on a lexer to break up each record into tokens. We use the open source software Flex~\cite{flex} as the lexer in our implementation. Accordingly, users need to write a Flex specification file tailored to their dataset in order to obtain a better tokenization scheme. We will compare against RecordBreaker in Section~\ref{sec_github_exp}.
\vspace{3pt}

 \noindent \textbf{Experiment Settings:} All experiments were conducted on a 64-bit Windows machine with 8-core Intel Xeon 3.40GHz CPU and 8GB RAM. All executions are single-threaded.

\subsection{Evaluation Criteria}\label{sec_exp_standard}

\begin{figure}[h]
	\vspace{-10pt}
	\begin{center}
		\includegraphics[width = 7cm]{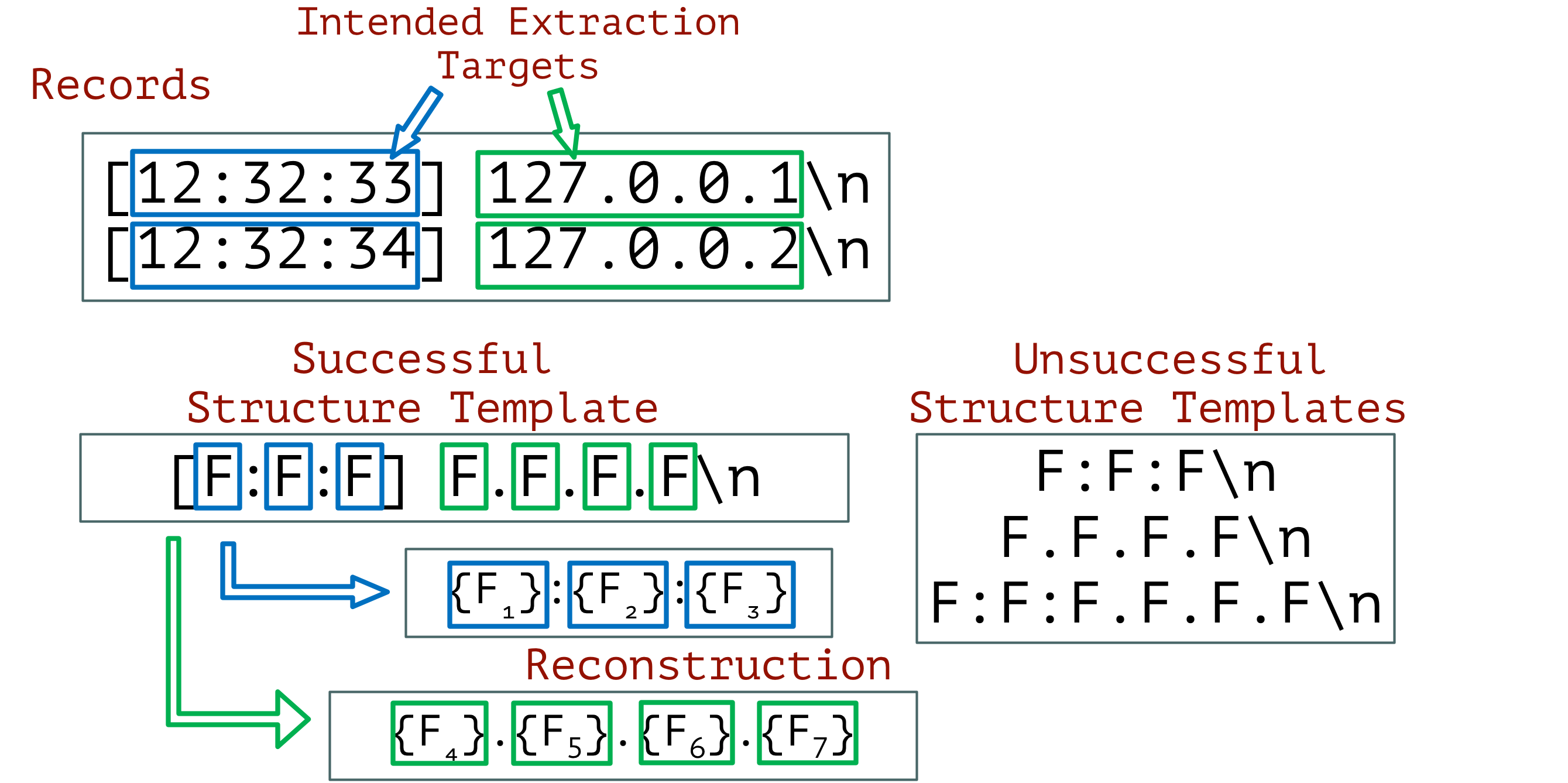}
	\end{center}
	\vspace{-10pt}
	\caption{Successful and Unsuccessful Extraction Examples}\label{fig_aggregation_script}
	\vspace{-15pt}	
\end{figure}

Recall that the structure extraction problem is not well-posed, and the validity of the extracted structure solely depends on the end-user. For many datasets, there are usually multiple structures that can potentially be deemed as valid. For example, the dataset \verb|[01:05:02]| \verb|192.168.0.1| has at least the following $4$ valid structure templates:
\begin{verbatim}
[F] F\n              [F] F.F.F.F\n
[F:F:F] F\n          [F:F:F] F.F.F.F\n
\end{verbatim}

\new{Thus, it is not possible to directly compare the extracted structure with a manually labeled structure. } In this paper, we define the \new{following} evaluation criteria: for each dataset, we first identify \new{several different types of records within the dataset, then identify} as many intended extraction targets as possible \new{for each type of record} (i.e., observable fields with potentially interesting information). The extraction is considered successful \new{{\em if both of the following two criteria are met: (a) all of the record boundaries and record types are correctly identified; (b) for each type of intended extraction target, we can select several fields from the structure template, such that all of the intended extraction targets (of this type) can be reconstructed by concatenating the selected fields from the corresponding record.} Figure~\ref{fig_aggregation_script} demonstrates an example successful extraction, in which we have two types of intended extraction targets (i.e., time and IP address), and \algname returns the structure template as shown in the middle of the figure. In this example, the extraction is considered successful because both types of intended extraction targets can be reconstructed by concatenating field values at specific positions for all extracted records. If, instead, the targets were extracted together, reconstructing them via concatenation would not be possible.}

A more rigorous version of the above evaluation criteria can be found in Section~\ref{sec_script_def} in appendix.

\subsection{Manually Collected Datasets}\label{sec_example_dataset}

The first $15$ datasets in this collection come from Fisher et al.'s work~\cite{fisher2008dirt}. Since Fisher's collection lacks large or complex datasets (i.e., datasets with multiple types of records or multi-line records), we also collected $10$ additional datasets from the internet (e.g., the stack exchange data dump~\cite{stackexchange}) as well as from our genomics collaborators. 
The sources and characteristics of the $25$ manually collected datasets can be found in Section~\ref{sec_source} in the appendix.

\stitle{Evaluation Goal.} The goal of the experiments in this section is to study various properties of \algname\footnote{\scriptsize \new{We do not compare with RecordBreaker in this section. RecordBreaker employs Fisher's algorithm~\cite{fisher2008dirt} and all $15$ datasets from Fisher's collection are used in this section. Therefore, RecordBreaker will likely perform very well on these datasets, and thus comparison on such datasets would not be objective and meaningful. We will however show that \algname can handle all $25$ datasets effectively.}}: in Section~\ref{sec_exp_accuracy}, we demonstrate the extraction accuracy; in Section~\ref{sec_exp_efficiency}, we study the efficiency of \algname under various settings. \techreport{Finally in Section~\ref{sec_exp_sensitivity}, we study the parameter sensitivity of \algname. }\paper{Parameter sensitivity experiments, i.e., the impact of $M$, $L$, and $\alpha$ on running time and accuracy
can be found in our extended technical report.}



\subsubsection{Extraction Accuracy}\label{sec_exp_accuracy}

We used \algname to extract structures from the $25$ datasets, and the extractions are successful for all $25$ datasets based on the evaluation standard in Section~\ref{sec_exp_standard}. \algname correctly identified the record boundaries for all $25$ datasets, without knowing the span of records and the position of noise blocks beforehand. For datasets with multiple types of records, \algname can also correctly identify the type of each record.
Based on these results, we conclude that {\em \algname is capable of extracting structure from a wide variety of datasets such that end-users could reconstruct any intended target field value using the extracted structures with little extra effort (in most cases no extra effort at all)}.





\subsubsection{Running Time}\label{sec_exp_efficiency}

We study the efficiency of \algname here. We first run \algname on the $25$ datasets using the default parameters to study the connection between the characteristics of datasets (size/structural complexity) and the running time. \techreport{Then, we vary the parameters to study their impact on the efficiency of \algname.}

\vspace{5pt}

\begin{figure}[h]
\vspace{-5pt}
	\centering
	\begin{subfigure}[b]{0.22\textwidth}
		\includegraphics[width=\textwidth]{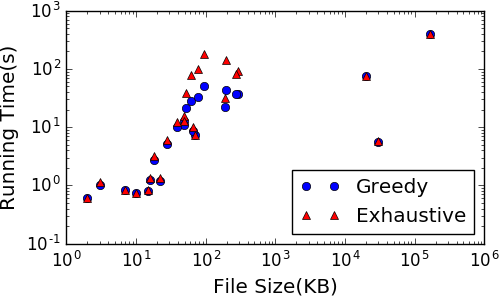}
		\caption{}\label{fig_running_time}
	\end{subfigure}
	~
	\begin{subfigure}[b]{0.22\textwidth}
		\includegraphics[width=\textwidth]{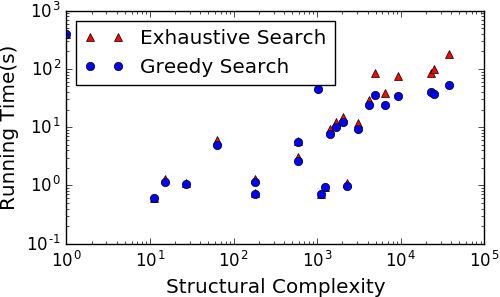}
		\caption{}\label{fig_running_time_struct}
	\end{subfigure}
	\vspace{-10pt}
	\caption{Running Time vs. (a) Dataset Size and (b) Structural Complexity. x axis in (b) is the number of structure templates with at least $10\%$ coverage.
	}
	\vspace{-10pt}	
\end{figure}

\noindent \textbf{Running Time vs. Dataset Size:} Figure~\ref{fig_running_time} depicts the impact of the size of the dataset on the running time of \algname (using either exhaustive search or greedy search). The running time on small datasets (less than $50$MB) is dominated by the generation and evaluation step. For these datasets, the average running time is $17$ seconds for greedy search and $37$ seconds for exhaustive search. It takes about $7$ minutes for \algname to process the largest dataset here (with size $167$MB), where the majority of the running time is spent on running the LL(1) parser~\cite{grune2007parsing} for the actual data extraction.
Note that the running time of the three major steps of \algname is not affected by dataset size for large datasets (as discussed in Section~\ref{sec_time_complexity}). As we can see in Figure~\ref{fig_running_time}, the extraction time is already dominated by the running time of LL(1) parser~\cite{grune2007parsing} (which is a necessary step for all structure extraction algorithms) even when the dataset is only moderately large (i.e., about $167$MB). Further, this step is easily parallelizable. Therefore, we conclude that \algname is efficient enough in practice.

\noindent \textbf{Running Time vs. Structural Complexity:}  Figure~\ref{fig_running_time_struct} depicts the impact of the structural complexity of the dataset on the running time of \algname. The structural complexity of datasets are characterized using the total number of structure templates with at least $10\%$ coverage. In general, it takes a longer time for \algname to extract datasets with higher structural complexity, and the efficiency benefits of greedy search is more significant on these datasets.


\techreport{\noindent \textbf{Running Time vs. Parameters:} Figure~\ref{fig_parameter_runtime} shows the impact of parameters on the running time of \algname (exhaustive search). Recall that $M$ is the number of remaining structure templates after pruning step. As we can see in the left figure, the value of $M$ directly affects the overall running time, and this effect is more significant for larger datasets. In the right figure, we can see that changing parameters $\alpha$ or $L$ also affect the efficiency of \algname. 

Note that if we evaluate all structure templates with at least $\alpha\%$ coverage (i.e., skipping the pruning step by setting $M = \infty$), the average running time would be longer than $6$ minutes even for small datasets. Therefore, it is necessary to use \apprxscore to prune out structure templates. 

\begin{figure}[h]
	\vspace{-5pt}
	\centering
	\begin{subfigure}[b]{0.22\textwidth}
		\includegraphics[width=\textwidth]{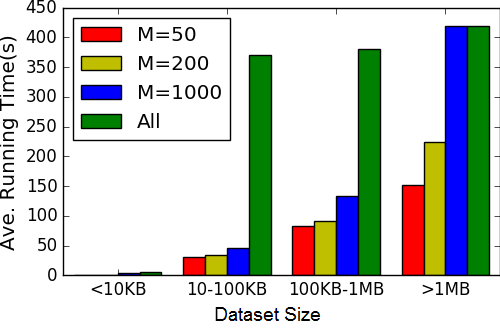}
	\end{subfigure}
	~
	\begin{subfigure}[b]{0.22\textwidth}
		\includegraphics[width=\textwidth]{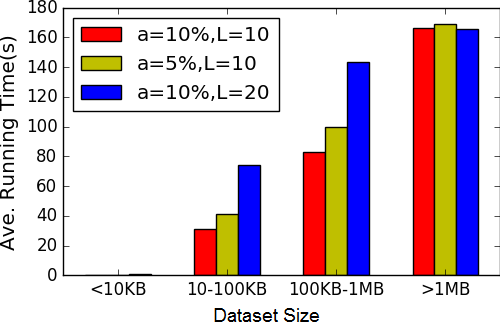}
	\end{subfigure}
	\vspace{-10pt}
	\caption{The impact of parameters on the running time.}\label{fig_parameter_runtime}
	\vspace{-15pt}	
\end{figure}
}

\techreport{
\subsubsection{Parameter Sensitivity}\label{sec_exp_sensitivity}
Since the extraction accuracy is not well suited for characterizing parameter impact (for most parameter configurations, the resulting structure would satisfy the requirements in Section~\ref{sec_exp_standard}), we use another metric to evaluate the impact of parameters: whether \algname can find the optimal structure template (i.e., the structure template with best \score, this is found by evaluating the \score of every structure template with at least $\alpha\%$ coverage). Figure~\ref{fig_parameter_percentage} shows the percentage of datasets in which \algname can find the optimal structure template on different parameter combinations. As we can see, \algname is very robust with respect to the parameter settings: for example, changing the value of parameter $M$ from $50$ to $1000$ only increased the likelihood of finding the optimal structure by about $10\%$. Figure~\ref{fig_parameter_percentage} also verifies the effectiveness of the \apprxscore in practice: for $40\%$ of the datasets, the optimal structure also has the best \apprxscore.

\begin{figure}[h]
	\vspace{-10pt}
	\begin{center}
		\includegraphics[width = 5cm]{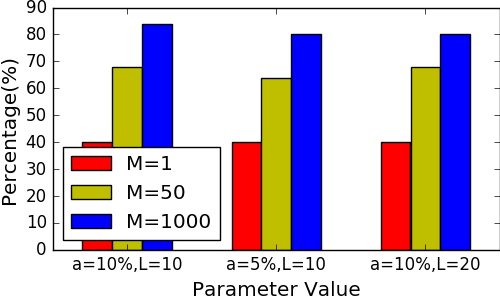}
	\end{center}
	\vspace{-10pt}
	\caption{The percentage of datasets in which \algname can find the optimal structure on different parameters}\label{fig_parameter_percentage}
		\vspace{-10pt}
\end{figure}

Note that it is not necessary for \algname to find the optimal structure, any structure that respects the criterion in Section~\ref{sec_exp_standard} is sufficient. The metric used in this section is solely for comparison purposes. Combining with the results in Section~\ref{sec_exp_efficiency}, we suggest using the following default parameter configuration in practice: $\alpha = 10\%$, $L = 10$, $M = 1000$.

}

\subsection{GitHub Datasets}\label{sec_github_exp}

GitHub contains a large quantity of log datasets generated by 
programmers across the world. 
We collected $100$\footnote{\scriptsize The scale is limited to $100$ since we have to manually inspect the datasets and the extraction results. \algname can be automatically applied to thousands of datasets without any problem. } of such datasets by {\em uniformly sampling} from the first $1000$ search results using the following three criteria: (a) files end with ``.log'' (b) with length greater than $20000$ (c) contains one of the following keywords\footnote{\scriptsize GitHub search function requires at least one search keyword, and we used multiple keywords to improve the diversity of our selection.} : ``db'', ``2016'', ``system'', ``query'', ``user''. The datasets are sampled using computer-generated random numbers and chosen before any follow-up analysis is conducted, so it represents an unbiased subset of the whole dataset. The characteristics of these datasets are discussed in Section~\ref{sec_github_characteristics}, and the experimental results are discussed in Section~\ref{sec_github_accuracy}. The $100$ sampled datasets constitute a new benchmark for structure extraction from log datasets, which will be released to public if this paper is accepted.

\stitle{Evaluation Goal.} The goal of the experiments in this section is to demonstrate the effectiveness of \algname on common log datasets ``in the wild''. In Section~\ref{sec_github_characteristics}, we study the characteristics of the log datasets in our sampled collection. In Section~\ref{sec_github_accuracy}, we evaluate the extraction accuracy of \algname and compare with RecordBreaker~\cite{recordbreaker}.


\subsubsection{Dataset Characteristics}\label{sec_github_characteristics}

The sampled datasets are categorized based on three criteria: 
\begin{denselist}
	\item {\em whether the dataset contains multi-line records}
	\item {\em whether the dataset consists of multiple types of records}
	\item {\em whether the dataset has any structure at all}
\end{denselist}  

\begin{figure}[h]
	\vspace{-10pt}
	\centering
	\begin{subfigure}[b]{0.22\textwidth}
		\includegraphics[width=\textwidth]{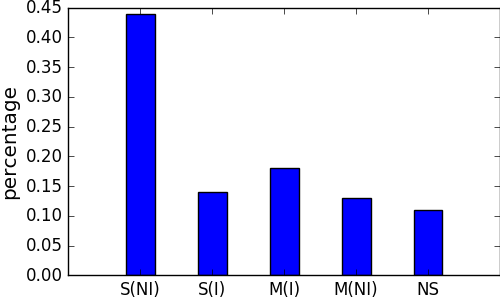}
		\caption{Characteristics}\label{fig_github_pie}
	\end{subfigure}
	~
	\begin{subfigure}[b]{0.22\textwidth}
		\includegraphics[width=\textwidth]{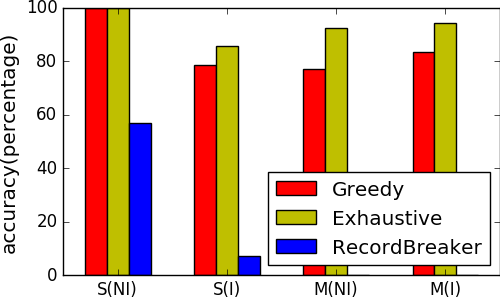}
		\caption{Accuracy}\label{fig_github_accuracy}
	\end{subfigure}
	\vspace{-10pt}
	\caption{GitHub Datasets: Characteristics and Accuracy}
	\vspace{-10pt}	
\end{figure}

\noindent There are five possible labels of datasets based on the above criteria, which are listed in Table~\ref{tbl_github_label}. The distribution of labels among the $100$ sampled log files is shown in Figure~\ref{fig_github_pie}. 

\begin{table}[h]
	\vspace{-10pt}

	\centering
	\footnotesize
	\begin{tabular}{|c|l|}
		\hline
		Label & Description \\
		\hline
		\hline
		S (Single-line) & Dataset consists of only single-line records. \\ 
		\hline
		M (Multi-line) & Dataset contains records spanning multiple lines\\
		\hline
		NI (Non-Interleaved) & Dataset consists of only one type of records.\\
		\hline
		I (Interleaved) & Dataset contains more than one types of records.\\
		\hline
		NS (No Structure) & Dataset has no structure or its structure does \\ 
			& not follow assumptions in Section~\ref{sec_assumption}.\\
		\hline
	\end{tabular}
	\caption{GitHub Dataset Labels}\label{tbl_github_label}
	\vspace{-15pt}
\end{table}

\noindent In the following, we discuss several findings from Figure~\ref{fig_github_pie}:

\begin{denselist}
	\item \noindent \textbf{Validity of Structural Assumptions:} 	
	89\% of datasets follow assumptions in Section~\ref{sec_assumption}, and 10\% of the datasets has no structure at all (nothing can be extracted from these datasets), only $1\%$ dataset have structure that cannot be described within the framework in Section~\ref{sec_assumption}. These statistics suggest that assumptions in Section~\ref{sec_assumption} are well-justified for log datasets.
	
	\item \noindent \textbf{Necessity for Multi-line Record Handling:}  
	31\% of datasets contains at least one type of record spanning multiple lines. The optimal structure in these datasets cannot be successfully extracted if the extraction system cannot handle multi-line records.
	
	\item \noindent \textbf{Necessity for Interleaved Records Handling:} 32\% of datasets contains more than one type of records. If the extraction system cannot recognize the existence of multiple types of records, only one type of record can be extracted (the rest will be regarded as noise), resulting information loss.
	
\end{denselist}





\subsubsection{Structure Extraction Accuracy}\label{sec_github_accuracy}

We applied \algname to extract structured information from GitHub datasets. Figure~\ref{fig_github_accuracy} shows extraction accuracy for different types of datasets (based on the standard in Section~\ref{sec_exp_standard}). Overall, \algname successfully extracted structure from $85$ datasets. The accuracy is $95.5$\% if we exclude datasets with no structure. 


As we can see in Figure~\ref{fig_github_accuracy}, \algname achieved $100$\% accuracy on single-line non-interleaved datasets, the simplest type of dataset. The accuracy of \algname for the other three types of datasets are $85.7\%$, $92.3\%$ and $94.4\%$ for exhaustive search, and $78.6\%$, $76.9\%$, $83.3\%$ for greedy search. Therefore, we conclude that \algname is effective for most of the log datasets in practice. We also identified major causes for inaccurate extractions, which can be found in Section~\ref{sec_cause} in the appendix.

Figure~\ref{fig_github_accuracy} also shows the extraction accuracy of RecordBreaker~\cite{recordbreaker} with default configurations and parameters for comparison. As we can see, RecordBreaker performs very poorly on log datasets with accuracy $56.8$\% and $7.1$\% on S(NI) and S(I) respectively and $0$\% on M(NI) and M(I), for a total of $29.2$\% accuracy,  which is not very surprising: RecordBreaker is originally designed for well-structured datasets, and cannot handle the noise-heavy log datasets very well. Furthermore, the resulting structure templates depend a lot on the Flex configurations and the tuning of two parameters in RecordBreaker (i.e., MaxMass and MinCoverage). This is because Flex configurations decide the quality of tokenization, while the other two parameters determine the datatype (i.e., {\em struct}, {\em array} or {\em union}) for a given list of records. However, there are no generic configurations or parameter values that work for all datasets, {\em which makes RecordBreaker less desirable in an unsupervised setting and incomparable to \algname.}

Figure~\ref{fig_github_pie} and Figure~\ref{fig_github_accuracy} also demonstrates why prior work such as RecordBreaker~\cite{recordbreaker} is not well-suited for extracting structure from log datasets: for any dataset containing multi-line records, the task of partitioning such dataset into collection of records is nontrivial (due to the presence of noise \& the fact that record span is unknown). From Figure~\ref{fig_github_pie}, we see that at least $31\%$\footnote{\scriptsize This number is an underestimate since Assumption~\ref{ass_tokenize} can also be violated in some datasets} of datasets cannot be handled by RecordBreaker~\cite{recordbreaker} as demonstrated by M(NI) and M(I) in Figure~\ref{fig_github_accuracy}.

\begin{figure*}[!t]
	\techreport{\vspace{-35pt}}
	\paper{\vspace{-55pt}}
	\begin{center}
		\includegraphics[width = 13cm]{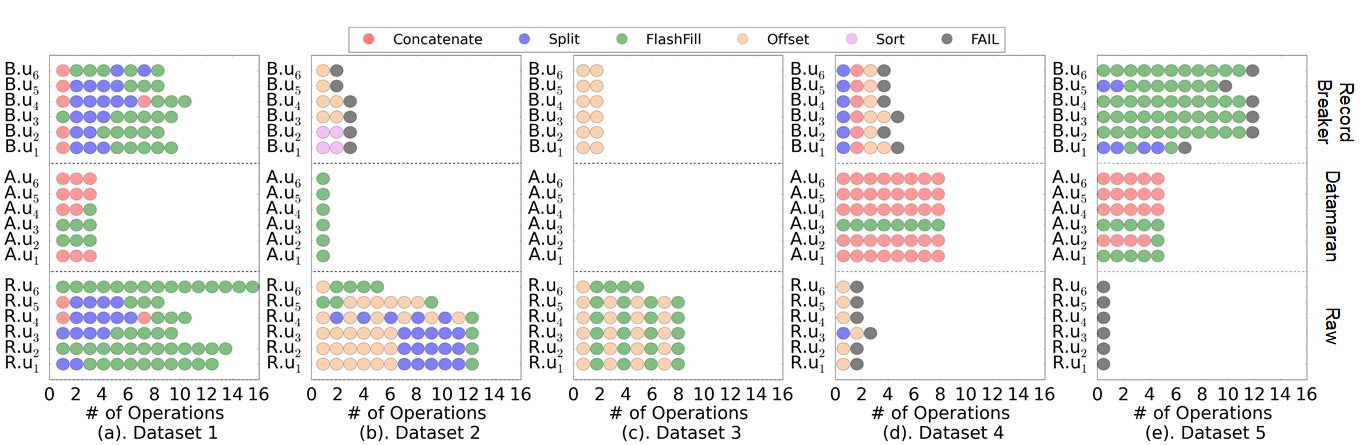}
	\end{center}
	\vspace{-10pt}
	\caption{Sequence of Operations for Transformation}\label{fig:user_study}
	\vspace{-10pt}
\end{figure*}

\section{User Evaluation} \label{sec_user_study}

To further evaluate the quality of the structure extracted by \algname, we conduct a user study on five representative log datasets, comparing our results against the raw datasets as well as the extracted results using RecordBreaker.

\paper{\vspace{-10pt}}

\subsection{Study Design}
Our user study simulates the following scenario, 
where a participant is presented with a log file, 
and they want to extract some information of interest, prior to analysis. 
One straightforward way to do so is to import the log file 
into a spreadsheet tool like Microsoft Excel, 
and then use Excel functionalities to extract this information. 
Alternatively, the participant can first use either \algname or 
RecordBreaker to extract the structure, 
and then refine the results using Excel to obtain the desired structure and filter out anything that is not of interest. 
We will compare these three methods (i.e., from the raw log file, from the result of \algname/RecordBreaker) in our user study.
In order to quantify the manual effort taken to reach the desired extraction result, 
we create a target extraction result based on our best judgement and 
use it as our gold standard. 
For each dataset, we show the raw log file as well as 
the extraction results of \algname and RecordBreaker to the participants, 
and ask them to transform each file into the target structure.

\paper{\vspace{-2pt}}
\stitle{Methodology.} The user study consists of three phases, in brief: 

{\em (1) Introduction phase:} We first show the participant an example of the raw file, extraction results from \algname and RecordBreaker, along with the target file, denoted as $R$, $A$, $B$ and $T$ respectively. Then, we introduce four popular Excel data wrangling functionalities that may be used for transforming those three files into the target file, {\em Concatenate, Split, FlashFill} and {\em Offset}.
{\em Concatenate} and {\em Split} are straightforward; {\em Flashfill} autocompletes columns from a few user examples~\cite{gulwani2011automating}; and {\em Offset} 
can be used to copy contents every $K$ rows while skipping the $(K-1)$ rows in-between. Overall, {\em Concatenate}, {\em Split}, and {\em FlashFill} are very easy to use, while {\em Offset} requires more thought and effort and is not very intuitive\paper{~\cite{DatamaranTechReport}}.

{\em (2) Quiz phase:} We present five folders to the participant, one for each dataset, where each folder contains the raw file ($R$), two extraction files ($A$ and $B$) and the target file ($T$). One dataset is a single-line record dataset while the other four are multi-line record datasets. For each dataset, the participant is asked to transform $R$, $A$, and $B$ into $T$ using the described functionalities in Excel, or any other functionalities they may be aware of. The whole process takes around one and half hours per participant.

{\em (3) Survey phase:} We conduct a survey to understand the participant's experience in structure extraction using the raw file $R$ and the two extraction files ($A$ and $B$).

\techreport{More specifically, in the introduction phase, we give an tutorial on the usage of four common data wrangling features in Excel: {\em Concatenate, Split, FlashFill} and {\em Offset} in Microsoft Excel. Let us first review the functionality of each operation.
{\em Concatenate} merges the strings from multiple cells into a combined string. 
Opposite to {\em Concatenate}, {\em Split} splits a string into multiple cells via delimeters.
{\em Offset} can be used to copy contents every $K$ rows while skipping the $(K-1)$ rows in-between. 
For example, {\em offset}$(B\$1, (row()-1)\times 5, 0, 1, 1)$ refers to the cell with $(row()-1)\times 5$ row offset and $0$ column offset from the reference cell $B\$1$, where $row()$ is the row id.
By specifying this formule, Excel extracts content every 5 rows and skips the 4 rows in between. In our user study, {\em Offset} helps reconstruct records spanning multiple rows.
Different from these cell-based operations, i.e., {\em Concatenate}, {\em Split}, and {\em Offset}, {\em FlashFill} is content-based.
It can automatically fill the data if it detects a pattern between input examples and the original data in Excel. 
That said, the functionalities of {\em FlashFill} can cover those provided by both {\em Concatenate} and {\em Split}. 
However, compared to {\em Split} which splits each column into multiple columns simultaneously, {\em FlashFill} can only fill in one column at a time.
Furthermore, {\em FlashFill} sometimes detects the wrong pattern, but by providing a few more examples, {\em FlashFill} can correct the mistake and provide the correct results.
Next, let us check the complexity for using each operation.
{\em Concatenate}, {\em Split} and {\em FlashFill} are very easy to use, while {\em Offset} requires more thought and effort in writing the formule since it involves the manipulations over multiple rows and is not very intuitive.

In the quiz phase, among the representative five datasets we present to the participants, one of them is a single-line dataset while the other four are multi-line datasets. Among the four multi-line datasets, two of them have a regular pattern, while the other two have noise. The raw dataset $R$, the extracted result using \algname $A$ and the target result $T$ is stored in a single file each, while there may be multiple files for the extracted results using RecordBreaker due to "union" structure type in their algorithm. We output the extraction results using RecordBreaker into multiple files if it recognizes multiple structures. 
}


\paper{\vspace{-2pt}}
\stitle{Participants.} We had six participants in our study, including five graduate students from Computer Science and one graduate student from Electrical and Computer Engineering. Four out of the six work with data very often (daily), one often (weekly) and one rarely (yearly or fewer). In addition, every participant has used spreadsheets and scripting language(s), like Python and Matlab, for data analysis, while two participants had also used business analytics tools like Tableau and Power BI.

\paper{\vspace{-5pt}}
\subsection{Result Analysis}
\noindent
\fbox{\begin{minipage}{26em} \small
    \textit{Summary:} We find {\em (a)} Starting from the extracted results using RecordBreaker and \algname, i.e., $A$ and $B$, helps the participant "fast-forward" to the desired target structure, compared to the raw file $R$. {\em (b)} The extracted result by \algname is already in a fine-grained clean format, requiring very simple operations, i.e., Concatenate or FlashFill, to concatenate the fine-grained results to get the target format $T$. {\em (c)} For multi-line datasets, it is hard to obtain the target information from both the raw file $R$ and the extracted result using RecordBreaker $B$, as evidenced by the {\bf failures} (black circles) in Figure~\ref{fig:user_study}. 
\end{minipage}}
\vspace{0.3mm}

For each dataset, we recorded the action sequences performed by each participant during the transformation. 
In total, there are $6\times 3 \times 5 = 90$ sequences, since we have six participants, three file types ($A$, $B$ and $R$), and five datasets. 
Each sequence is depicted by a horizontal line in Figure~\ref{fig:user_study}, where each colored circle denotes a specific operation\footnote{We ignore the simple operations like Delete, Copy, Paste.} performed by the participant, as shown in the legend. 
The x-axis is the operation's index in the sequence, and y-axis shows the participant id and the file type.
For instance, $R.u_1$ refers to the first participant ($u_1$) and the task is to transform the raw file ($R$) into the target file.

As shown in Figure~\ref{fig:user_study}, participants took more operations to transform the raw file (if no failure occurred) as opposed to extracted files using \algname and RecordBreaker. 
This verifies the usefulness of automated extraction tools.
Furthermore, participants always took the least number of steps to reach the target file $T$ when using \algname, with no failure.
On the contrary, they were often unable to transform the raw file $R$ and the extracted file using RecordBreaker $B$, as shown in Figure~\ref{fig:user_study}(b,d-e).
This occurred mostly when the records span multiple-lines and when the dataset is noisy. 
Next, we will discuss the findings for each dataset briefly. More details can be found \techreport{in Section~\ref{ssec_drilldown}}\paper{in our technical report~\cite{DatamaranTechReport}}.

Dataset 1 is a single-line dataset, and the extraction results of both RecordBreaker and \algname are much better structured than the raw file\techreport{ as shown in Figure~\ref{fig:dataset_1}}. Compared to $R$ and $B$, $A$ took the smallest number of steps in order to be transformed to $T$, as illustrated in Figure~\ref{fig:user_study}(a).
When it comes to multi-line record datasets, i.e., datasets 2-5, \algname exhibits a much more substantial advantage over RecordBreaker and the raw file. 
First, when there is noise or incomplete records in the dataset (dataset 4 and 5), participants needed to either manually filter the incomplete records one by one, or write some sophisticated code to remove the noise and reconstruct the records. This step is often laborious or hard to implement. 
Second, RecordBreaker treats each single line as a record unit, and would recognize each line as a different structure, which are then stored into different files. 
Hence, the participants often found themselves losing context for reconstructing the records when each record spanned multiple files.
As a consequence, participants often failed to transform $B$ and $R$ into $T$ after some trials, as shown by the black circles in Figure~\ref{fig:user_study}(b,d-e).
Due to the context missing in $B$, participants could only figure out that they failed to reconstruct the rows after a number of operations, as illustrated in Figure~\ref{fig:user_study}(e). 

\paper{\vspace{-15pt}}
\subsection{Survey and Interview}
\vspace{1mm}
\noindent
\fbox{\begin{minipage}{26em} \small
    \textit{Summary:} All participants ranked the extracted results by \algname ($A$) easiest to use, and the raw file ($R$) most difficult to use. 
    This is mostly because the structure in the raw file is unclear, while \algname provides a very clear structure.
\end{minipage}}
\vspace{0.3mm}

Most participants (5/6) reported that $A$ (\algname) is very easy to use, requiring only merging (i.e., {\em Concatenate} and {\em FlashFill}) and deleting operations most of the time.
But some participants also complained that $A$ still requires a bunch of manual work, like repeating {\em Concatenate}. This is because the extraction results of \algname is of a very fine-grained nature, 
The large number of repeating operations of {\em Concatenate} or {\em FlashFill} is captured in Figure~\ref{fig:user_study}(d). 
On the other hand, all participants (6/6) complained that the raw file is hard to begin with, since it looks messy and is difficult to find the pattern inside.
In addition, participants were not satisfied with the extracted results by RecordBreaker, since they were annoyed by the multi-file and multi-line merge operations like {\em Offset}. 
On average, participants rated the difficulty of performing transformation from $A$, $B$, and $R$ to $T$ as $1.8$, $7.8$, and $9.3$ respectively, where $1$ indicates the easiest and $10$ indicates the hardest.

In particular, one participant ($u_4$) said the following---"For $A$, it is ready to use, involving mostly merge and delete operations. 
For $B$, there is lots of extra operations. It's hard to carefully use Offset to merge lines and merging across rows could be painful and error prone.
For $R$, it is impossible to do manually. I prefer to write code, but need to make sure the code is bug free." 
Another participant ($u_6$) said the following---"No major difficulty for $A$. Each row corresponds to exactly one record. 
For $B$, there is information lost during processing, hard (impossible?) to join disparate partially processed items together.
$R$ requires significant manual effort to identify anomalous records before automatic techniques can be applied to put data in structured format."
There is also some limitations identified for \algname ($A$).
One participant ($u_1$) said the following---"For $A$, it only involves single file operators, easier to track, but still a lot of manual work.
For $B$, it requires cross file operations, difficult to track, and sometimes you end up choosing sub-optimal operations.
For $R$, it is unstructured, need to create tuple using Offset first, most laborious among the three."

From the user study, we conclude that \algname has better extraction results than RecordBreaker, and both tools are a better starting point than the raw file. 

\stitle{Limitations.} Since our user study is limited to the comparison between two automated structure extraction tools, i.e., \algname and RecordBreaker, and supervised extraction starting from the raw file using techniques like FlashFill, it remains to be seen whether unsupervised tools can perform as well as other more advanced supervised extraction tools. 
Also, the many concatenate operations (e.g., assembling IP addresses from fragments) can be tedious. 
For such domain-specific datatypes. \algname should be enhanced with type awareness (e.g., for phone numbers, IPs, URLs).

\paper{\vspace{-10pt}}
\section{Related Work}\label{sec:related}
Our work is related to the vast bodies of work on general information extraction,
as well as the more limited work on log dataset extraction, and string transformation.
Other related work can be found in Section~\ref{app_related}.

\stitle{Unsupervised HTML Wrapper Induction.}
A few papers attempt to extract from HTML pages directly, without requiring 
any training examples~\cite{arasu2003extracting,crescenzi01roadrunner,sleiman2014trinity,sleiman2013tex}.
All of these papers rely on repetitiveness within a page, or the redundancy
across similar pages to separate the content from the template.
The rules that are inferred are strongly dependent on the HTML DOM tree structure;
in our case, we do not have the luxury of HTML tags to distinguish
between records or fields.

\stitle{Extraction from Web Documents.}
There has been some work on extraction from other forms of documents, or portions of Web documents,
typically leveraging example concepts~\cite{senellart2008automatic} or 
a knowledge-base~\cite{agichtein2004mining,cortez2011joint,zhao2008exploiting,li2008regular}
to extract entities and attributes from text files.

List extraction, i.e., extraction from lists on the web
is another area that has seen some work~\cite{elmeleegy2009harvesting,chjwww02,zhai2005web,gupta2009answering,machanavajjhala2011collective}.
Some of these papers require both the eventual relational schema as well as candidate examples to be provided~\cite{gupta2009answering,machanavajjhala2011collective}. Some papers
attempt fully-automated list extraction~\cite{elmeleegy2009harvesting,chjwww02,zhai2005web}.
These papers make the crucial assumption of each record corresponding to a single list item, making it easy
to extract the boundaries of the records. 
\techreport{Our space of datasets---log files---do not admit any such assumption.}

\stitle{Log Dataset Extraction and Transformation.} Wrangler~\cite{kandel2011wrangler,DBLP:conf/uist/GuoKHH11} supports the interactive specification
of log dataset cleaning operations, drawing from the transformations in Raman et al.~\cite{raman2001potter}.
Instead of operator specification, other work relies on user-provided input-output examples~\cite{jin2017foofah,DBLP:conf/popl/Gulwani11,DBLP:journals/cacm/GulwaniHS12,le2014flashextract,barowy2015flashrelate} to transform one semi-structured dataset to another. 
\techreport{In our case, we do not require any intervention from the user.}
\new{The PADS project~\cite{fisher2008dirt} relies on a user-provided chunker and tokenizer to identify the boundaries of records/field values, while RecordBreaker is a line-by-line unsupervised implementation, with a fixed lexer configuration which makes it inflexible 
for real log datasets. Recent work by Raza and Gulwani~\cite{raza2017automated} describe
an automatic text extraction DSL for single-line extraction, generalizing to both web-pages and text documents.}

\new{Other work clusters event logs~\cite{vaarandi2004breadth, makanju2009clustering} by 
treating the lines of the log dataset as data points and assigning them to clusters. Compared to our work, these papers do not attempt to identify the structure within records, and they do not consider the possibility of multi-line records.}



\paper{\vspace{-10pt}}

\section{Conclusions}\label{sec_conclusion}

We presented \algname, a completely unsupervised automatic structure extraction tool specifically tailored towards log datasets.
\techreport{We formally defined the structure extraction problem as an optimization problem, where we are given a \scoreF that reflects human judgment, and we search for the best structure template that optimizes this \scoreF.
\algname algorithm consists of three major steps: the generation step searches for structure templates satisfying the minimum coverage threshold assumption; the pruning step orders them using an \apprxscoreF; the evaluation step evaluates the structure templates remaining after the pruning step, and further refines them to achieve even better score;
We experimentally evaluated \algname on a collection of representative datasets and a large collection of log files crawled from GitHub. }
The experimental results demonstrate that \algname can efficiently and correctly extract structures from all representative datasets and $95.5\%$ of the GitHub datasets, and is robust with respect to parameter choices, while RecordBreaker can only extract $29.2\%$ from the same dataset collection.
\new{Our user study further demonstrates that \algname, in addition to automatically extracting from log datasets, provides a valuable starting point for data analysis: all participants (6/6) preferred \algname to RecordBreaker and the raw file.}

\smallskip
\noindent {\bf Acknowledgments.} We acknowledge support from NSF grants IIS-1513407, IIS-1633755, IIS-1733878 and IIS-1652750 and funds from Adobe, Google, Toyota, and the Siebel Energy Institute.

\bibliographystyle{abbrv}
\bibliography{dbbib}

\vspace{-10pt}

\section{Appendix}

\subsection{Other Algorithmic Details}\label{sec_detail}

Here we discuss some additional algorithmic and implementation details of \algname that were not covered in the main body of the paper. \paper{Pseudocode can be found in our technical report\cite{DatamaranTechReport}. }


\vspace{5pt}

\noindent \textbf{Variants of Generation Step.} We implemented two searching procedures in \algname 
for finding the optimal {\em RT-CharSet}. Both searching procedures require {\em RT-CharSet-Candidate}, the set of characters that can potentially be in {\em RT-CharSet}, as an input. 

Suppose there are $c$ different characters in {\em RT-CharSet-Candidate} that appeared in the dataset. The exhaustive search would enumerate all $2^c$ subsets. On the other hand, the greedy search procedure would only enumerate $O(c^2)$ of them. The greedy search procedure operates in the following way: initially, {\em RT-CharSet} is set to be empty; then in each step, one of the characters in {\em RT-CharSet-Candidate} is added to {\em RT-CharSet}; the decision for choosing which character to add is made greedily by choosing the character generating the structure template with highest \apprxscore (as defined in Section~\ref{sec_pruning}). 
\paper{An example illustrating the two searching procedures can be found in our technical report~\cite{DatamaranTechReport}.}

\techreport{
The following example helps illustrate the two searching procedures. 
Consider a dataset with the following structure template: \Verb|[F:F:F] F(F,F)|.}

\techreport{There are $7$ special characters in total: \Verb|[|\Verb|]|\Verb|:|\Verb|(|\Verb|)|\Verb|,|(space character). Thus, the exhaustive search would enumerate $128$ possible subsets for this example. As for the greedy search, it starts from the empty set and gradually adds new characters into it:}

\techreport{
\begin{denselist}
	\item in the first step, it enumerates all the subsets containing only one character, and computes the corresponding structure templates (i.e., invoking steps 2-5).
	\item it then decides which subset to proceed based on which one 
	has the structure template with the highest \apprxscore (for this example, it is ``\Verb|F:F:F|'').
	\item then in the second step, it enumerates all $6$ subsets consisting of the character `:' and one additional character.
	\item this procedure repeats until either the subset is full 
	or we can no longer find any structure template with at least $\alpha\%$ coverage.
\end{denselist} 
}

\techreport{
It is easy to see that, for this example, the maximum number of subsets that the greedy search 
would have enumerated is 29 (also counting the empty subset here). On the other hand, the exhaustive search would have enumerated $128$ subsets. Note that if the field values do not contain any special characters in {\em RT-CharSet-Candidate}, then the correct {\em RT-CharSet} would contain all characters in {\em RT-CharSet-Candidate} that appeared in the dataset. In this case, the greedy search procedure is guaranteed to find the correct {\em RT-CharSet} since it will always consider the full subset at the end of the searching procedure.
}



\vspace{5pt}

\noindent \textbf{Extracting Record Template From Instantiated Record.} The non-overlapping assumption (Assumption~\ref{ass_main}) states that there exists two disjoint sets of characters $A$ and $B$, such that for any instantiated record $R$, {\em RT-CharSet}$(R)$ (i.e., the record template character set) is a subset of $A$, and {\em F-CharSet}$(R)$ (i.e., the field value character set) is a subset of $B$. By this assumption, the record template can be uniquely extracted from any of its instantiated records given the value of $A$ and $B$. 
For example, if $A = \{$ \Verb|',', '\n'| $\}$, then the instantiated record \Verb|1,2,3,45,6,78,9,a,bc,d\n|
can be transformed into the record template \Verb|F,F,F,F,F,F,F,F,F,F\n| by replacing characters not in $A$ with the field placeholder.

\vspace{5pt}

\noindent \textbf{Reducing Record Templates to Structure Templates.} We identify the corresponding minimum structure template that can generate each extracted record template. This is achieved by repeatedly reducing repeated patterns into array regular expressions. For example, the record template \Verb|F,F,F,F,F| \Verb|,F,F\n| is reduced into the structure template \Verb|(F,)*F\n|. 
If there are conflicting reduction steps (i.e., reduction steps that cannot be performed simultaneously), we choose one of them arbitrarily. The reduction process only \new{guarantees that we find a minimal structure template (i.e., a structure template that cannot be reduced further), which means that not all instantiated records are reduced back to the same structure template. As a result, the coverage estimate during the generation step is an underestimate. However, in our experiments, the initial coverage estimate is usually still well above the $\alpha\%$ threshold, thereby not affecting the correctness of the generation step.}  

\vspace{5pt}

\noindent \textbf{Pruning Using Hash-Table.} We store all of the structure templates in a hash-table, and maintain the total coverage of all structure templates associated with each hash-bin. For all hash-bins with less than $\alpha \%$ total coverage, the associated structure templates are discarded. 

\techreport{
\vspace{5pt}

The pseudocode of the generation step can be found in Algorithm~\ref{alg_generation}. Two searching procedures correspond to function {\em GreedySearch} and {\em ExhaustiveSearch} respectively. The function {\em GenST} finds structure templates with at least $\alpha$\% coverage given the value of {\em RT-CharSet}.

}

\vspace{5pt}

\noindent \textbf{Handling Multiple Structure Templates.} In the cases where there are more than one 
type of record in the dataset, we repeat the entire structure detection process (Generation-Pruning-Evaluation) for multiple times. 
After each iteration, we retrieve the parts of the dataset that are not explained by the previous structure. These parts are concatenated together, and we run the entire procedure on it again.

\vspace{5pt}

\noindent \textbf{Sampling Technique.} In the actual implementation of \algname, sampling is used 
instead of simply scanning through the entire dataset in both the generation and evaluation step. 
For large datasets, scanning the whole dataset during these steps \new{is not feasible: the total number of whole dataset scans is equal to the number of {\em RT-CharSet}s enumerated in the generation step plus $M$ in the evaluation step}. Our sampling implementation is cache-aware: we sample several large chunks of data and concatenate them in the memory. 
Both the generation step and the evaluation steps are performed on the concatenated chunks instead.

\subsection{Default Regularity Score Function}\label{sssec_score}

We implemented a simple default \scoreF based on the minimum 
description length principle~\cite{barron1998minimum}: 
we design a record generation procedure from the structure template, 
and the \score is \new{equal to the total amount of information needed for describing 
all the instantiated records using the structure template}, plus the additional information needed to describe the noise blocks. 
For completeness, we describe the details of this score function in the following.
Describing the record using the structure template is straightforward 
given Assumption~\ref{ass_form}:
\begin{itemize}
	\item For arrays, we describe the number of repetitions, then describe each repetition individually.
	\item For structs, we describe each component individually.
	\item For fields, the description scheme depends on its value type.
\end{itemize}


For the field value description, we associate each \new{field in the structure template} with one of the following four 
value-types: enumerated type, integer, real number, or string.
The description schemes for field values depend on the data-type---which can be 
determined by analyzing the field values in the group; 
the details of these schemes are listed as follows: 
\begin{itemize}
	\item The enumerated type fields are described using $\lceil \log_2 n\_value \rceil$ bits, where $n\_value$ is the total number of unique values.
	\item The integer fields are described using $\lceil \log_2 (max\_value - min\_value + 1) \rceil$ bits, where $max\_value$ and $min\_value$ are the upper bound and lower bound of the field value, which can be determined by scanning through the dataset.
	\item The real number fields are described using $\lceil \log_2 [(max\_value - min\_value) \times 10^{exp} + 1] \rceil$ bits, where $max\_value$ and $min\_value$ are the same as above, and $exp$ is the maximum number of digits after the decimal point.
	\item The string fields are described directly using $(len(s) + 1) \times 8$ bits, where $len(s)$ is the length of the field value. The $+1$ term is to include the end-of-string \Verb|'\0'| character, and each character needs $8$ bits to describe. 
\end{itemize}

\noindent Using the description schemes above, the total description length can be computed as $ D(dataset) = len(ST) \times 8 + 32 + m + \sum_{i=1}^m$ $D(block_i)$.
The first $len(ST) \times 8$ bits describe the the structure template, and the next $32 + m$ bits describe
the total number of blocks in the dataset and whether each block is a noise block or a record. $D(block_i)$ is the description length of $i$th block: for noise blocks, it is simply the block length times $8$; for record blocks, we compute its description length accordingly.

\techreport{The pseudocode for computing the description score can be found in Algorithm~\ref{alg_eval}, 
with the following $3$ steps:
\begin{enumerate}
	\item extract all the instantiated records from the dataset.
	\item estimate the data-type parameters from the extracted records.
	\item compute the description length using the formulae above.
\end{enumerate}}

\subsection{Formal Evaluation Standard}\label{sec_script_def}

\new{In order to formalize our evaluation standard, we consider both the relational dataset extracted from \algname\footnote{For RecordBreaker, it is also possible to convert the extracted result into relational format, and therefore the evaluation standard also applies.} (the procedure of converting extracted results into relational format is described in Section~\ref{sec_ass_form}) and a relational dataset containing only the intended extraction targets. {\em We say the extraction is successful if it is possible to convert the extracted relational dataset into the target relational dataset via a sequence of the following relational operations:}
	\begin{denselist}
		\item \textbf{Concat$(R, C_1, C_2)$:} Create a new column in $R$. For each tuple $t$ in $R$, the new entry value is equal to the concatenation of the corresponding entries in column $C_1$ and $C_2$.
		\item \textbf{GroupConcat$(R_1, R_2, FK, C)$:} Create a new column in $R_1$. For each tuple $t$ in $R_1$, the new entry value is equal to the concatenation of entries in column $C$ of tuples in $R_2$ with foreign-key column $FK$ referencing $t$ (i.e., $C$ and $FK$ are columns of $R_2$, and $FK$ is a foreign-key column referencing $R_1$).
		\item \textbf{Trim$(R, C, pre, suf)$:} Remove the first $pre$ characters and the last $suf$ characters of all entries in column $C$ of relation $R$ (i.e., $pre$ and $suf$ are constant numbers).
		\item \textbf{Append$(R, C, pre\_str, suf\_str)$:} Add $pre\_str$ to the beginning and $suf\_str$ to the end of all entries in column $C$ of relation $R$ (i.e., $pre\_str$ and $suf\_str$ are constant strings).
		\item \textbf{DeleteCol$(R, C)$:} Delete column $C$ of relation $R$.
		\item \textbf{DeleteTable$(R)$:} Delete relation $R$.
	\end{denselist}
	\noindent In other words, we consider the extraction successful if the target relational dataset can be reconstructed by merging/removing some columns in the extracted relational dataset. Intuitively, this is only possible if (a) the fields are well-aligned within each column of the extracted relational dataset (i.e., they are of the same data type); and (b) the extracted relational dataset has more fine-grained splitting of fields compared to the intended extraction format. Note that we do not allow splitting columns here, otherwise even the trivial extraction result specifying the whole record as a single field would be considered successful.
}

\subsection{Causes for Inaccurate Extraction}\label{sec_cause}
Here we describe the causes for inaccurate extraction for GitHub log datasets (Section~\ref{sec_github_exp}). There are $4$ log files where even the exhaustive search version of \algname failed to find a valid structure. In the following, we list the two causes for these inaccurate extractions, and discuss the potential ways to address them.

\vspace{5pt}
\noindent \textbf{Fail to recognize ``long'' records:} The maximum range of records is set to be $10$ lines during the experiments. In some datasets, there are some extremely ``long'' records that exceeds this limit. If we increase the range limit, the efficiency of \algname would suffer. As the records in practice can be arbitrarily long, we are still unaware of methods that can completely solve this problem.

\vspace{5pt}
\noindent \textbf{The greedy approach for interleaved datasets:} In \algname, we handle interleaved datasets by repeatedly applying the algorithm on the dataset. However, this greedy procedure does not always find the correct structure for interleaved dataset. Instead, sometimes we would find structure templates with characteristics of multiple types of records. The following example illustrates this phenomenon. Suppose we have two types of records with templates:
\begin{verbatim}
F: F F F\n            F: F F F F F F\n
\end{verbatim}
\algname could potentially settle on the wrong structure template \Verb|F: (F )*F\n|, when this generic structure template has a lower \score compared to the two correct record templates.

\subsection{Sources and Characteristics of Manually Collected Datasets}\label{sec_source}

Table~\ref{tbl_dataset_source} lists the sources and characteristics of \techreport{
the $25$ manually collected datasets\footnote{For crash log datasets, there are two valid structures with max record span $1$ and $3$ respectively}. The first $15$ datasets are from Fisher et al.'s paper~\cite{fisher2008dirt} (marked with ``*'' in Table~\ref{tbl_dataset_source}).}\paper{the last 10 datasets that we use for our evaluation, in addition to the $15$ datasets from Fisher et al.'s paper~\cite{fisher2008dirt}---the maximum file size among the Fisher datasets is 0.3MB, with most of them being $\sim$0.02MB, with typically only 1 record type (for all datasets except one), and typically 1 for record span. Details can be found in the technical report.}

\begin{table}[h]
	\footnotesize
	\centering
	\begin{tabular}{|c|c|c|c|}
		\hline
		Data source & File size(MB) & \# of rec. types & Max rec. span \\
		\hline
		\hline
\techreport{*transaction records & 0.07 & 1 & 1 \\
		\hline
		*comma-sep records & 0.02 & 1 & 1 \\
		\hline 
		*web server log & 0.29 & 1 & 1 \\
		\hline 
		*log file of Mac ASL & 0.28 & 1 & 1 \\
		\hline
		*Mac OS boot log & 0.02 & 1 & 1 \\
		\hline
		*crash log & 0.05 & 1 & 1(3) \\
		\hline
		*crash log (modified in~\cite{fisher2008dirt}) & 0.05 & 1 & 1(3) \\
		\hline
		*ls -l output & 0.01 & 1 & 1 \\
		\hline
		*netstat output & 0.01 & 2 & 1 \\
		\hline
		*printer logs &  0.02 & 1 & 1 \\
		\hline
		*personal income records & 0.01 & 1 & 1 \\
		\hline
		*US railroad info & 0.01 & 1 & 1 \\
		\hline
		*application log & 0.06 & 1 & 1 \\
		\hline
		*LoginWindow server log & 0.05 & 1 & 1 \\
		\hline
		*pkg install log & 0.02 & 1 & 1 \\
		\hline}
		Thailand district info & 0.19 & 1 & 8 \\
		\hline
		stackexchange xml data & 20 & 1 & 1 \\
		\hline
		vcf genetic format & 167.4 & 1 & 1 \\
		\hline
		fastq genetic format & 29.9 & 1 & 4 \\
		\hline
		blog xml data & 0.06 & 1 & 10 \\
		\hline
		log file (1) & 0.03 & 2 & 9 \\
		\hline
		log file (2) & 0.01 & 1 & 3 \\
		\hline
		log file (3) & 0.19 & 2 & 1 \\
		\hline
		log file (4) & 0.07 & 2 & 10 \\
		\hline
		log file (5) & 0.09 & 1 & 4 \\
		\hline
	\end{tabular}
	\caption{Sources and characteristics of manually collected datasets. }\label{tbl_dataset_source}
	\vspace{-10pt}
\end{table}

\begin{figure*}[t]
\centering
	\begin{subfigure}{0.4\textwidth}
		\centering
		\includegraphics[width=\linewidth]{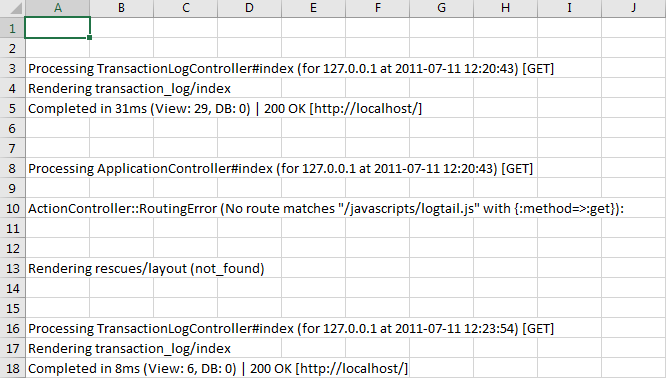}
		\vspace{-5mm}
		\caption{Raw File ($R$)}
		\label{fig:5_r}
	\end{subfigure}
	\hspace{6mm}
	\begin{subfigure}{.45\textwidth}
		\centering
		\includegraphics[width=\linewidth]{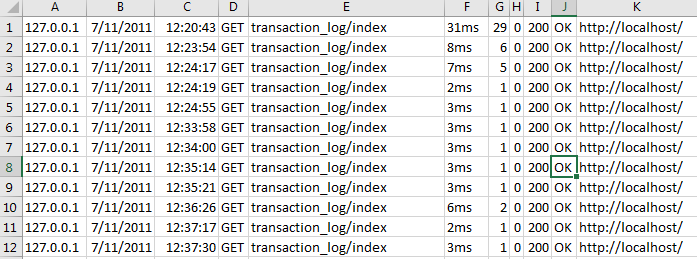}
		\vspace{5mm}
		\caption{Target File ($T$)}
		\label{fig:5_t}
	\end{subfigure}
	\begin{subfigure}{0.8\textwidth}
		\centering
		\includegraphics[width=\linewidth]{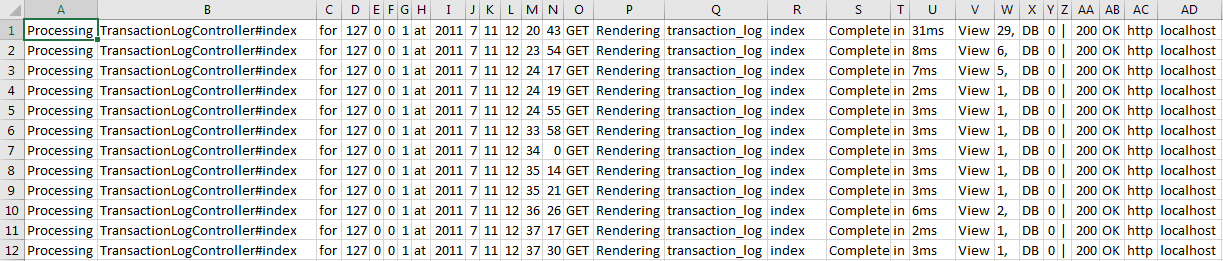}
		\vspace{-5mm}
		\caption{Extraction Result by \algname ($A$)}
		\label{fig:5_a}
	\end{subfigure}
	\hspace{5mm}
	\begin{subfigure}{.39\textwidth}
		\centering
		\includegraphics[width=\linewidth]{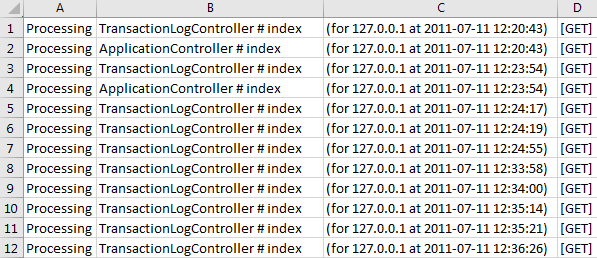}
		\vspace{-5mm}
		\caption{$B.1$}
		\label{fig:5_b1}
	\end{subfigure}
	\hspace{1mm}
	\begin{subfigure}{.3\textwidth}
		\centering
		\includegraphics[width=\linewidth]{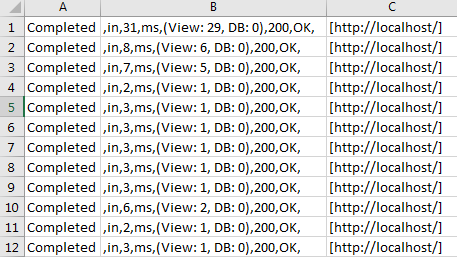}
		\vspace{-5mm}
		\caption{$B.2$}
		\label{fig:5_b2}
	\end{subfigure}
	\hspace{1mm}
	\begin{subfigure}{.16\textwidth}
		\centering
		\includegraphics[width=\linewidth]{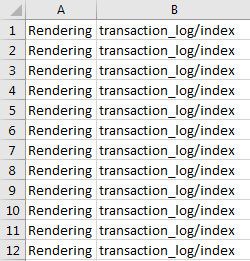}
		\vspace{-5mm}
		\caption{$B.3$}
		\label{fig:5_b3}
	\end{subfigure}
	\vspace{-2mm}
	\caption{Multi-Line Dataset with Noise (Dataset 5)}
	\label{fig:dataset_5}
	\vspace{-5mm}
\end{figure*}

\subsection{Proof of Theorem~\ref{thm_performance}}\label{sec_thm_prf}

\begin{proof}
	First of all, condition (b) ensures that $ST_0$ can be found during the generation step. Then, using condition (a), we can ensure that $ST_0$ to be the top structure template during the pruning step. Finally, condition (c) ensures that $ST_0$ will be chosen during the evaluation step. Combining all arguments, we can see that \algname is guaranteed to return $ST_0$ as the optimal structure template.
\end{proof}

\new{
\subsection{Drill Down on User Evaluation}\label{ssec_drilldown}
In our user study, we evaluated three different types of datasets: a single-line record dataset (dataset 1), multi-line record dataset with a regular pattern (dataset 2-3), and multi-line record dataset with noise (dataset 4-5). 
\paper{In the following, we only drill down on dataset 5, as a representative of the most complicated case, i.e., a noisy multi-line record log file. Additional drill down analyses can be found in our technical report~\cite{DatamaranTechReport}}.\techreport{In the following, we drill down on dataset 1, 3, and 5, as representatives of different dataset types.}
}

\techreport{
\begin{figure*}[t]
\centering
	\vspace{2mm}
	\begin{subfigure}{0.36\textwidth}
		\centering
		\includegraphics[width=\linewidth]{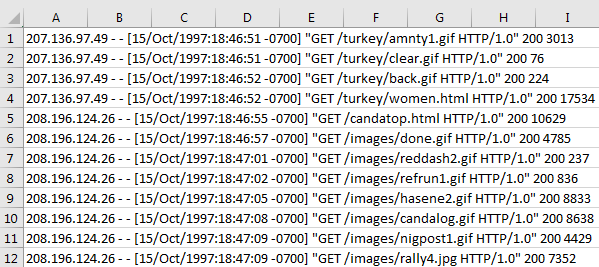}
		\vspace{-5mm}
		\caption{Raw File ($R$)}
		\label{fig:1_r}
	\end{subfigure}
	\hspace{5mm}
	\begin{subfigure}{.4\textwidth}
		\centering
		\includegraphics[width=\linewidth]{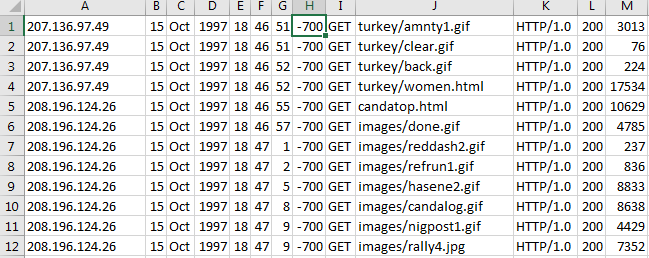}
		\vspace{-5mm}
		\caption{Target File ($T$)}
		\label{fig:1_t}
	\end{subfigure}
	\begin{subfigure}{0.4\textwidth}
		\centering
		\includegraphics[width=\linewidth]{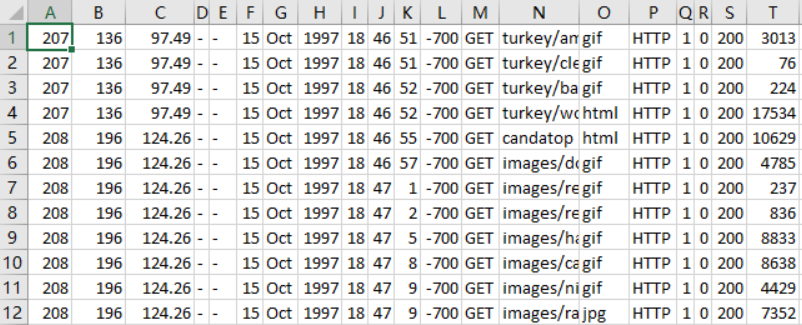}
		\vspace{-5mm}
		\caption{Extraction Result by \algname ($A$)}
		\label{fig:1_a}
	\end{subfigure}
	\hspace{5mm}
	\begin{subfigure}{.4\textwidth}
		\centering
		\includegraphics[width=\linewidth]{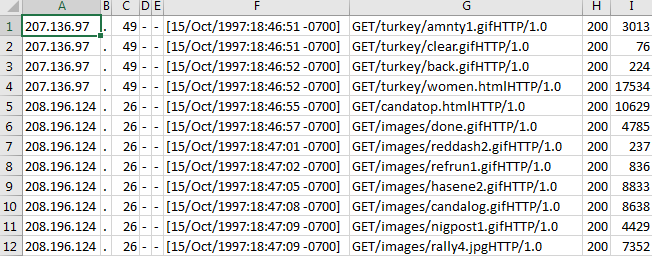}
		\vspace{-5mm}
		\caption{Extraction Result by RecordBreaker ($B$)}
		\label{fig:1_b}
	\end{subfigure}
	\vspace{-2mm}
	\caption{Single-Line Dataset (Dataset 1)}
	\label{fig:dataset_1}
	\vspace{-3mm}
\end{figure*}

\stitle{Single-Line Dataset.} Dataset 1 is a web server log, with one record per line.
Figure~\ref{fig:dataset_1} depicts the original file ($R$), the target file ($T$), the extraction result using \algname ($A$), and the extraction result using RecordBreaker ($B$) for dataset 1.
We can see that both $A$ and $B$ are in a clean form, and closer to the target file compared to the raw file. 
In comparison with $B$, each field in $A$ is of a fine-grained granularity.
In the following, we will illustrate how participants transform $R$, $A$ and $B$ into $T$ as depicted in Figure~\ref{fig:user_study}, respectively.

\begin{itemize}
\item {\em From $A$ to $T$.} Most columns are already perfect, except column A, J and K in Figure~\ref{fig:dataset_1}(b). 
Participants either performs Concatenate or FlashFill to obtain those columns.
For all participants, the total number of operations is 3, though with varying number of FlashFill and Concatenate.
\item {\em From $B$ to $T$.} participants either perform Concatenate or FlashFill on column A and B in Figure~\ref{fig:dataset_1}(d) to form column A in Figure~\ref{fig:dataset_1}(b). 
Next, participants all perform a sequence of Split with different delimeters on column F in Figure~\ref{fig:dataset_1}(d) to obtain B-H in Figure~\ref{fig:dataset_1}(b).
Last, participants use FlashFill to derive column I-L in Figure~\ref{fig:dataset_1}(b) from column G in Figure~\ref{fig:dataset_1}(d).
The total number of operations is 9. 
\item {\em From $R$ to $T$.} To extract column A in Figure~\ref{fig:dataset_1}(b), participants either performs Concatenate or FlashFill.
For the other columns, participants can either use Split or FlashFill.
But compared to using Split, FlashFill may involve more steps. For instance, Split by space on $R$ can successfully extract column A, L and M in $T$; 
while with FlashFill, it requires to repeat three times to extract those three columns. 
In particular, one participant ($u_2$) simply perform FlashFill to extract the targeted 13 columns in Figure~\ref{fig:dataset_1}(b).  
The average total number of operations is 10.
\end{itemize}

\begin{figure*}[t]
\centering
	\begin{subfigure}{0.24\textwidth}
		\centering
		\includegraphics[width=\linewidth]{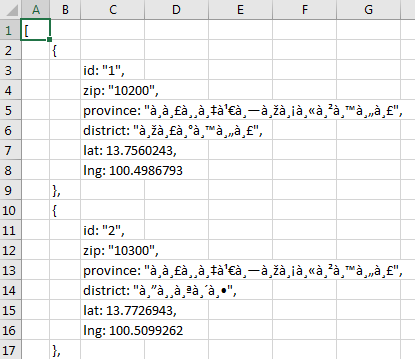}
		\vspace{-5mm}
		\caption{Raw File ($R$)}
		\label{fig:3_r}
	\end{subfigure}
	\hspace{35mm}
	\begin{subfigure}{.18\textwidth}
		\centering
		\includegraphics[width=\linewidth]{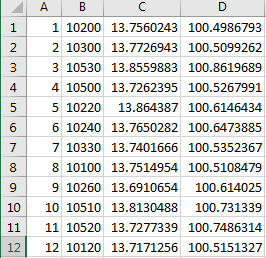}
		\vspace{-1mm}
		\caption{Target File ($T$)}
		\label{fig:3_t}
	\end{subfigure}
	\begin{subfigure}{0.4\textwidth}
		\centering
		\includegraphics[width=\linewidth]{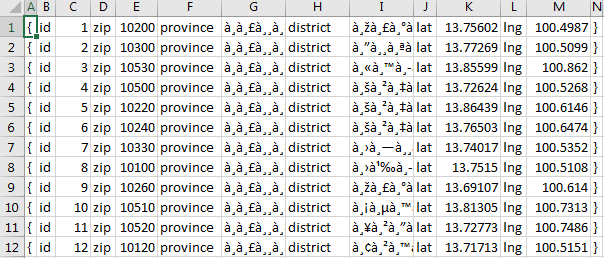}
		\vspace{-5mm}
		\caption{Extraction Result by \algname ($A$)}
		\label{fig:3_a}
	\end{subfigure}
	\hspace{15mm}
	\begin{subfigure}{.083\textwidth}
		\centering
		\includegraphics[width=\linewidth]{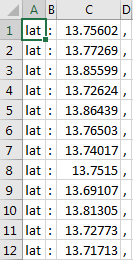}
		\vspace{-5mm}
		\caption{$B.1$}
		\label{fig:3_b1}
	\end{subfigure}
	\hspace{1mm}
	\begin{subfigure}{.072\textwidth}
		\centering
		\includegraphics[width=\linewidth]{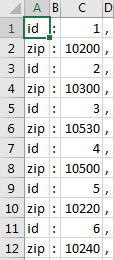}
		\vspace{-5mm}
		\caption{$B.2$}
		\label{fig:3_b2}
	\end{subfigure}
	\hspace{1mm}
	\begin{subfigure}{.082\textwidth}
		\centering
		\includegraphics[width=\linewidth]{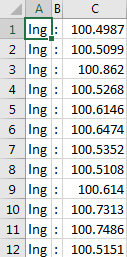}
		\vspace{-5mm}
		\caption{$B.3$}
		\label{fig:3_b3}
	\end{subfigure}
	\vspace{-2mm}
	\caption{Multi-Line Dataset with Regular Pattern (Dataset 3)}
	\label{fig:dataset_3}
	\vspace{-5mm}
\end{figure*}

\stitle{Multi-Line Dataset with Regular Pattern.}
Dataset 3 is a clean JSON file recording Thailand district information, with one record spanning multiple lines.
Figure~\ref{fig:dataset_3} depicts the original file ($R$), the target file ($T$), the extraction result using \algname ($A$), and the extraction result using RecordBreaker ($B$) for dataset 3.
The raw file is in JSON format, with each record spanning multiple lines.
\algname successfully identify the correct schema and group multi-lines into one record.
On the contrary, RecordBreaker treat each record as three different structure due to the difference in Column C and D in Figure~\ref{fig:dataset_3}(d-f).
It is easy to combine Figure~\ref{fig:dataset_3}(d) and (f), while Figure~\ref{fig:dataset_3}(e) contains two different fields.
participants need to use the more advanced functionality, i.e., Offset, in Excel to manually align the attribute and reconstruct the rows. 
Similar thing must be used to transform the raw file into the target file.
Note that for most participants, this is the very first time to know and use Offset functionality.
In the following, we will illustrate how participants transform $R$, $A$ and $B$ into $T$ as depicted in Figure~\ref{fig:user_study}, respectively.

\begin{itemize}
\item {\em From $A$ to $T$.} Nothing needs to be done here, except copying and pasting the desired 4 columns A-D in Figure~\ref{fig:dataset_3}(b).
\item {\em From $B$ to $T$.} First, participants perform Offset two times in Figure~\ref{fig:dataset_3}(e) to separate the value of id from the value of zip, and store them as two separate columns. 
Next, we can copy and paste the content from Figure~\ref{fig:dataset_3}(e-f) to Figure~\ref{fig:dataset_3}(d). Thus, the total number of operations is 2.
\item {\em From $R$ to $T$.} Since each target row spans multiple lines in the raw file. Offset is first used to create column A-D in Figure~\ref{fig:dataset_3}(b). 
Most participants perform Offset four times, while one participant ($u_6$) was able to use one unified Offset formule to extract all four columns A-D.
Correspondingly, FlashFill is used to extract the value inside each field. The average total number of operations is 8. 
\end{itemize}

}

\stitle{Multi-Line Dataset with Noise.}
\new{
Dataset 5 is a real log dataset crawled from GitHub, with each record spanning multiple lines.
Noise and incomplete records exist in this dataset.
Figure~\ref{fig:dataset_5} depicts the original file ($R$), the target file ($T$), the extraction result using \algname ($A$), and the extraction result using RecordBreaker ($B$) for dataset 5.
As readers may have already noticed, the raw file has no regular patterns. More specifically, Line 3-5 in Figure~\ref{fig:dataset_5}(a) is a block unit forming one record, while Line 8, 10 and 13 are noise data or incomplete records.
As a consequence, it is impossible to reconstruct the records via {\em Offset} in Excel, since there is no regular pattern in the raw file. 
Similarly, RecordBreaker also fails to handle such noisy datasets, because it cannot filter incomplete records or noise from the desired target.
For instance, Line 2 in Figure~\ref{fig:dataset_5}(d) corresponds to the noise data, i.e., Line 8 in Figure~\ref{fig:dataset_5}(a).
On the contrary, \algname works well with multi-line noisy datasets, and can successfully extract fine-grained attributes from the raw dataset.
In the following, we will illustrate how participants transformed $R$, $A$ and $B$ into $T$ as depicted in Figure~\ref{fig:user_study}, respectively.
\begin{denselist}
\item {\em From $A$ to $T$.} Participants simply used Concatenate or FlashFill to merge columns in Figure~\ref{fig:dataset_5}(c) into column A-C, E and K in Figure~\ref{fig:dataset_5}(b). 
For instance, by combining column D-G in Figure~\ref{fig:dataset_5}(c), we can obtain column A in Figure~\ref{fig:dataset_5}(b). The total number of operations is 5.
\item {\em From $B$ to $T$.} Participants first tried FlashFill and Split to extract the target information, but then they found that the partial contents in Figure~\ref{fig:dataset_5}(d), e.g., line 2, belong to the incomplete records, and it is hard to tell the noisy data from the target ones. Thus, participants failed to transform $B$ into $T$.
\item {\em From $R$ to $T$.} After looking at the raw file, participants all found it impossible to convert $R$ to $T$ via Excel. This is due to the existence of noise and incomplete records.
\end{denselist}
}

\techreport{
\begin{algorithm}[h]
	\scriptsize
	\caption{The Generation Step}\label{alg_generation}
	\begin{algorithmic}[5]
		\Function{GenST}{$char\_set$}
		\State $n \leftarrow \text{Total Number of Lines}$
		\For{$i \leftarrow 1$ to $n$}
		\For{$j \leftarrow i + 1$ to $i + L$}
		\State $left\_boundary \leftarrow i$
		\State $right\_boundary \leftarrow j$
		\State $r \leftarrow \text{ExtractRecord}(left\_boundary, right\_boundary)$
		\State $rt \leftarrow \text{ExtractRecordTemplate}(r, char\_set)$
		\State $st \leftarrow \text{GenerateStructureTemplate}(rt)$
		\State $k \leftarrow \text{ComputeHashKey}(st)$
		\State $cov(k) \leftarrow cov(k) + \text{length}(r)$
		\State $st\_set(k) \leftarrow st\_set(k) \cup \{r\}$
		\EndFor
		\EndFor
		\State Find all hash keys with more than $\alpha$\% coverage.
		\State \Return the associated structure templates.
		\EndFunction
		\Function{ExhaustiveSearch}{$char\_candidates$}
		\State $ST\_set \leftarrow \emptyset$ 		
		\For{$char\_set \subseteq char\_candidates$}
		\State $ST\_set \leftarrow ST\_set \cup \text{GenST}(char\_set)$
		\EndFor
		\State \Return $ST\_set$
		\EndFunction
		\Function{GreedySearch}{$char\_candidates$}
		\State $char\_set \leftarrow \emptyset$
		\State $ST\_set \leftarrow \emptyset$ 
		\Repeat
		\State $new\_best\_char\_set \leftarrow \emptyset$
		\State $best\_f \leftarrow 0$
		\For{$c \in char\_candidates \setminus char\_set$}
		\State $new\_char\_set \leftarrow char\_set + c$
		\State $new\_ST\_set \leftarrow \text{GenST}(new\_char\_set) $
		\State $ST\_set \leftarrow ST\_set \cup new\_ST\_set$
		\For{$st \in new\_ST\_set$}
		\If{$\text{AssScore}(st) > best\_score$}
		\State $best\_score \leftarrow \text{AssScore}(st)$
		\State $new\_best\_char\_set \leftarrow new\_char\_set$
		\EndIf
		\EndFor
		\EndFor
		\State $char\_set \leftarrow new\_best\_char\_set$
		\Until {\text{no structure template has at least $\alpha$\% coverage}} 
		\State \Return $ST\_set$
		\EndFunction
	\end{algorithmic}
\end{algorithm}

\begin{algorithm}[h]
	\scriptsize
	\caption{The Evaluation Step}
	\label{alg_eval}
	\begin{algorithmic}[5]
		\Function{EvalST}{$ST$}
		\State $(RecordBlocks, NoiseBlocks) \leftarrow \text{ParseData}(ST)$
		\State Determine the data types of field values
		\State Learn the distributional parameters
		\State $TotalDL \leftarrow len(ST) \times 8 + 32 + NumBlocks$
		\For{$record \in RecordBlocks$}
		\State $RT \leftarrow \text{GetRecordTemplate}(record)$
		\State $TotalDL \leftarrow TotalDL + D(RT|ST)$
		\State $TotalDL \leftarrow TotalDL + D(record|RT)$
		\EndFor
		\For{$block \in NoiseBlocks$}
		\State $TotalDL \leftarrow TotalDL + len(block) \times 8$
		\EndFor
		\State \Return $TotalDL$
		\EndFunction
		\Function{RefineST}{$ST$}
		\Repeat
		\State $ST' \leftarrow \text{UnfoldArray}(ST)$
		\If{$\text{EvalST}(ST) > \text{EvalST}(ST')$}
		\State $ST \leftarrow ST'$
		\EndIf 
		\Until{$ST$ cannot be further unfolded}
		\State $ST \leftarrow \text{ShiftStructure}(ST)$
		\State \Return $ST$
		\EndFunction
	\end{algorithmic}
\end{algorithm}

}

\subsection{Other Related Work} \label{app_related}
We now briefly mention other related work that we didn't cover in the main body of the paper.

\stitle{Example-Driven HTML Wrapper Induction.}
There has been a long line of work on inducing or learning a ``wrapper'' to extract
content from HTML pages, e.g., \cite{dalvi09robust,freitag2000boosted,han01wrapping,hsu98generating,muslea98stalker,vertex,kushmerick1997wrapper}.
The majority of these papers crucially rely on both the web-page structure in the form of the DOM,
as well as on text (e.g., extract the piece of text immediately following ``Price:'').
Examples are provided in the form of entities that belong to the concept class
that are to be extracted, or in the form of explicit annotations (e.g., this location 
contains an item of interest to be extracted). 
Often, the eventual relational schema is known in advance. 
Some papers do not rely on the HTML structure, opting instead to
use NLP~\cite{etzioni2004web,agichtein2000snowball}.
In our case, we do not require any seed entities or annotations.

\stitle{Extracting Structure From Other Media.} \new{There is 
	work~\cite{antunes2011reverse, cui2007discoverer, Cohen2011, Spitkovsky2011} on
	extracting structure from other types of media (i.e., other than text-formatted
	log datasets). The extraction strategies adopted by these papers crucially rely on 
	characteristics of the target dataset type. For instance, in security research
	~\cite{antunes2011reverse, cui2007discoverer,wang2011inferring, bossert2014towards,wang2012semantics, whalen2010hidden},	the network traces consist of continuous 
	communication between server and client, best modeled
	as a deterministic state machine (i.e., messages between server and client represent transitions in the global state),
	and reliant on indicators that signal the start of a new message, e.g., the presence of an IP address;
	in either case, the record boundaries are clear. 
    On the other hand, in the field of natural language processing~\cite{Cohen2011, Spitkovsky2011}, 
    the structure is usually restricted to local context (i.e., within each sentence), 
    and can be captured using probabilistic language models. 
    In particular, Cohen et al.~\cite{Cohen2011} employs language models from other languages to 
    learn the structure of a new language, while Spitkovsky et al.~\cite{Spitkovsky2011} uses clustering
    based on local context (neighboring words to a given word) to infer dependency structures to inform a sentence parser, 
    where parsing is
    delimited based on periods. 
    In our case, the fundamental characteristics of log datasets are captured in Definition~\ref{def_dataset}, 
    and our whole extraction strategy revolves around this definition.
}

\end{document}